\documentclass[twoside,11pt]{article}

\usepackage{graphicx,eepic,epic,epsfig,amsmath,latexsym,amssymb,psfrag}

\usepackage{theorem}
\newtheorem{definition}{Definition}[section]
\newtheorem{proposition}[definition]{Proposition}
\newtheorem{lemma}[definition]{Lemma}

\newtheorem{theorem}[definition]{Theorem}
\newtheorem{corollary}[definition]{Corollary}

\def\squareforqed{\hbox{\rlap{$\sqcap$}$\sqcup$}}
\def\qed{\ifmmode\squareforqed\else{\unskip\nobreak\hfil
\penalty50\hskip1em\null\nobreak\hfil\squareforqed
\parfillskip=0pt\finalhyphendemerits=0\endgraf}\fi}
\def\endenv{\ifmmode\;\else{\unskip\nobreak\hfil
\penalty50\hskip1em\null\nobreak\hfil\;
\parfillskip=0pt\finalhyphendemerits=0\endgraf}\fi}
\newenvironment{proof}{\noindent \textbf{{Proof~} }}{\qed}
\newenvironment{remark}{\noindent \textbf{{Remark~}}}{\qed}
\newenvironment{example}{\noindent \textbf{{Example~}}}{\qed}

\newcommand{\exampleTitle}[1]{\textbf{(#1)}}

\mathchardef\ordinarycolon\mathcode`\:
\mathcode`\:=\string"8000
\def\vcentcolon{\mathrel{\mathop\ordinarycolon}}
\begingroup \catcode`\:=\active
  \lowercase{\endgroup
  \let :\vcentcolon
  }

\newcommand{\captionfonts}{\small}
\makeatletter  % Allow the use of @ in command names
\long\def\@makecaption#1#2{%
  \vskip\abovecaptionskip
  \sbox\@tempboxa{{\captionfonts \textbf{#1}\; #2}}%
  \ifdim \wd\@tempboxa >\hsize
    {\captionfonts \textbf{#1}\; #2\par}
  \else
    \hbox to\hsize{\hfil\box\@tempboxa\hfil}%
  \fi
  \vskip\belowcaptionskip}
\makeatother   % Cancel the effect of \makeatletter

\newcommand{\nc}{\newcommand}
\nc{\rnc}{\renewcommand}
\nc{\beq}{\begin{equation}}
\nc{\eeq}{{\end{equation}}}
\nc{\beqa}{\begin{eqnarray}}
\nc{\eeqa}{\end{eqnarray}}
\nc{\lbar}[1]{\overline{#1}}
\nc{\bra}[1]{\langle#1|}
\nc{\ket}[1]{|#1\rangle}
\nc{\ketbra}[2]{|#1\rangle\!\langle#2|}
\nc{\braket}[2]{\langle#1|#2\rangle}
\nc{\proj}[1]{| #1\rangle\!\langle #1 |}
\nc{\avg}[1]{\langle#1\rangle}
\rnc{\max}{\operatorname{max}}
\nc{\Rank}{\operatorname{Rank}}
\nc{\smfrac}[2]{\mbox{$\frac{#1}{#2}$}}
\nc{\Tr}{\operatorname{Tr}}
\nc{\ox}{\otimes}
\nc{\dg}{\dagger}
\nc{\dn}{\downarrow}
\nc{\cA}{{\cal A}}
\nc{\cB}{{\cal B}}
\nc{\cC}{{\cal C}}
\nc{\cD}{{\cal D}}
\nc{\cE}{{\cal E}}
\nc{\cF}{{\cal F}}
\nc{\cG}{{\cal G}}
\nc{\cH}{{\cal H}}
\nc{\cI}{{\cal I}}
\nc{\cJ}{{\cal J}}
\nc{\cK}{{\cal K}}
\nc{\cL}{{\cal L}}
\nc{\cP}{{\cal P}}
\nc{\cR}{{\cal R}}
\nc{\cS}{{\cal S}}
\nc{\cT}{{\cal T}}
\nc{\cX}{{\cal X}}
\nc{\cZ}{{\cal Z}}
\nc{\csupp}{{\operatorname{csupp}}}
\nc{\qsupp}{{\operatorname{qsupp}}}
\nc{\rar}{\rightarrow}
\nc{\lrar}{\longrightarrow}

\def\a{\alpha}

\def\d{\delta}
\def\e{\epsilon}

\def\t{\theta}

\def\l{\lambda}

\def\p{\pi}
\def\r{\rho}
\def\s{\sigma}

\def\ph{\varphi}
\def\c{\chi}

\def\o{\omega}

\def\G{\Gamma}

\def\Ph{\Phi}

\nc{\RR}{{{\mathbb R}}}
\nc{\CC}{{{\mathbb C}}}
\nc{\FF}{{{\mathbb F}}}
\nc{\NN}{{{\mathbb N}}}
\nc{\ZZ}{{{\mathbb Z}}}
\nc{\PP}{{{\mathbb P}}}
\nc{\QQ}{{{\mathbb Q}}}
\nc{\UU}{{{\mathbb U}}}
\nc{\EE}{{{\mathbb E}}}

\begin{document}

\title{{\bf Trading quantum for classical resources\protect\\ in quantum data compression}}
\author{Patrick Hayden$^*$, Richard Jozsa$^\dg$ and Andreas Winter$^\dg${\vspace{.3cm}}\\
    {\small $^*$Institute for Quantum Information, Caltech, Pasadena, CA 91125 USA}\\
    {\small Email: {\tt patrick@cs.caltech.edu}}{\vspace{.2cm}}\\
    {\small $^\dagger$Department of Computer Science, University of Bristol}\\
    {\small Merchant Venturers Building, Bristol BS8 1UB, U.K.}\\
    {\small Email: {\tt richard@cs.bris.ac.uk, winter@cs.bris.ac.uk}}
}

\date{\today}

\maketitle

\begin{abstract}
We study the visible compression of a source $\cE = \{
\ket{\ph_i},p_i \}$ of pure quantum signal states, or, more
formally, the minimal resources per signal required to represent
arbitrarily long strings of signals with arbitrarily high
fidelity, when the compressor is given the identity of the input
state sequence as classical information. According to the quantum
source coding theorem, the optimal quantum rate is the von Neumann
entropy $S(\cE)$ qubits per signal.

We develop a refinement of this theorem in order to analyze the
situation in which the states are coded into classical and quantum
bits that are quantified separately. This leads to a trade--off
curve $Q^*(R)$ where $Q^*(R)$ qubits per signal is the optimal
quantum rate for a given classical rate of $R$ bits per signal.

Our main result is an explicit characterization of this trade--off
function by a simple formula in terms of only single signal,
perfect fidelity encodings of the source. We give a thorough
discussion of many further mathematical properties of our formula,
including an analysis of its behavior for group covariant sources
and a generalization to sources with continuously parameterized
states. We also show that our result leads to a number of
corollaries characterizing the trade--off between information gain
and state disturbance for quantum sources. In addition, we
indicate how our techniques also provide a solution to the
so--called remote state preparation problem. Finally, we develop a
probability--free version of our main result which may be
interpreted as an answer to the question: ``How many classical
bits does a qubit cost?''  This theorem provides a type of dual to
Holevo's theorem, insofar as the latter characterizes the cost of
coding classical bits into qubits.

\end{abstract}
\vfill\pagebreak

\tableofcontents

\section{Introduction} \label{sec:introduction}
When the term ``quantum information'' was first coined, it would
have been hard to predict how thorough and fruitful the analogy
between quantum mechanics and classical information theory would
ultimately prove to be.  The general approach, characterized by
the treatment of quantum states as resources to be manipulated,
has yielded a promising collection of applications, ranging from
unconditionally secure cryptographic protocols
\cite{bennett:brassard,lo:security,mayers} to quantum algorithms
\cite{deutsch:jozsa,shor,simon}. Moreover, the analogy, which was
initially unavoidably vague, has gradually been filled in by a
diverse variety of rigorous theorems describing achievable limits
to the manipulation of quantum states, such as the
characterization of the classical information capacity of quantum
sources \cite{holcap,schum:west}, of the optimal strategies for
entanglement concentration and dilution
\cite{entanglementConcentration} and many more. One of the pivotal
results of the emerging theory is the quantum source coding
theorem \cite{BFJS96,JozsaS94,schumacher}, demonstrating that for
the task of compressing quantum states, the von Neumann entropy
plays a role directly analogous to the Shannon entropy of
classical information theory.  Indeed, the quantum theorem
subsumes the classical one as the special case in which all the
quantum states to be compressed are mutually orthogonal.

A quantum source (or ensemble) $\cE = \{ \ket{\ph_i},p_i \}$ is
defined by a set of pure quantum signal (or ``letter'') states
$\ket{\ph_i}$ with given prior probabilities $p_i$ (cf. below for
precise definitions of these and other terms used in the
introduction). In this paper we will study the so--called {\em
visible} compression of $\cE$.  More specifically, we wish to
characterize the minimal resources per signal that are necessary
and sufficient to represent arbitrarily long strings of signals
with arbitrarily high fidelity, when the compressor is given the
identity of the input state sequence as {\em classical}
information (as the sequence of labels $i_1 \ldots i_n$ rather
than the quantum states $\ket{\ph_{i_1}}\ldots \ket{\ph_{i_n}}$
themselves, for example). According to the quantum source coding
theorem the optimal {\em quantum} rate in this scenario is the von
Neumann entropy $S(\cE)$ qubits per signal. We will develop a
refinement of this theorem in which the states are coded into
classical and quantum bits which are quantified {\em separately}.
This leads to a trade--off curve $Q^*(R)$ where $Q^*(R)$ qubits
per signal is the optimal quantum rate that suffices for a given
classical rate $R$ bits per signal. The quantum source coding
theorem implies that $Q^*(0)=S(\cE)$ and evidently we also have
$Q^*(H(p))=0$ where $H(p)$ is the Shannon entropy of the prior
distribution of the source. (By standard classical compression,
the compressor can represent the full information of the input
sequence in $H(p)$ classical bits per signal.) Thus the trade--off
curve extends between the limits $0\leq R\leq H(p)$.

There are various reasons why we might wish to maintain a
separation between classical and quantum resources in an encoding
\cite{BarnumHJW01}. On a purely practical level it seems to be far
easier to manufacture classical storage and communication devices
than it is to make quantum ones. But perhaps the primary reason is
conceptual: classical and quantum information have quite different
fundamental characters, with classical information exhibiting
special properties not shared by quantum information in general.
For example classical information is robust compared to quantum
information -- it may be readily stabilized and corrected by
repeated measurement that would destroy quantum information. Also,
unlike quantum information, it may be cloned or copied. These and
other singular properties indicate that for many purposes it may
be useful to regard classical information as a separate resource,
distinct from quantum information. Classical information is
sometimes formally regarded as a special case of quantum
information {\em viz.} the quantum information of a fixed set of
orthogonal states. While this characterization is useful for
formal analyses, it is unsatisfactory conceptually because it
relies on the essentially non-physical infinite
precision of orthogonality.  It is, therefore, perhaps better to
view classical information as a separate resource.

Exploring the trade--off possibilities between the two resources
will lead to a better understanding of the interrelation of these
concepts and the nature of quantum information itself. If bits can
always be represented as qubits (and indeed, by Holevo's
information bound~\cite{Holevo73}, at least one qubit per bit is
necessary and sufficient), what are the limitations on
representing qubits as bits? Under what conditions is it possible
at all? If there is a penalty to be paid, how large is it?  In
this paper we will give answers to these questions.

Our main result is a simple characterization of the trade--off
function $Q^*(R)$ which may be paraphrased as follows. Given the
ensemble $\cE=\{ \ket{\ph_i},p_i \}$ comprising $m$ states
$\ket{\ph_i}$ we consider decompositions of $\cE$ into at most
$(m+1)$ ensembles $\cE_j$ with associated probabilities $q_j$ i.e.
the ensembles $\cE_j = \{ \ket{\ph_i}, q(i|j)\}$ have the same
states as $\cE$ and their union $\bigcup_j q_j \cE_j$ reproduces
$\cE$. This is equivalent to the condition
\begin{equation}
p_i = \sum_j q(i|j) q_j
\end{equation}
on the chosen probabilities $q_j$ and $q(i|j)$ defining the
decomposition. Let $\overline{S}= \sum_j q_j S(\cE_j)$ be the
average von Neumann entropy of any such decomposition and let
$H(i:j)$ be the classical mutual information of the joint
distribution  $q(i,j)$.  For any $R$ let $\overline{S}_{\rm min}
(R)$ be the least average von Neumann entropy over all
decompositions that have $H(i:j)=R$. Then we will prove that the
trade--off function is given by $Q^*(R)=\overline{S}_{\rm min}
(R)$.

The prescription of a decomposition $\cE =\bigcup_j q_j \cE_j $
may be equivalently given in terms of a visible encoding map $E$
of the states of $\cE$:
\begin{equation}
 E(i) = \proj{\ph_i}\ox \sum_j p(j|i) \proj{j}.
\end{equation}
Here $p(j|i)$ are chosen freely subject only to the condition that
$H(i:j)=R$ and the previous probability distributions are
constructed as $q_j=\sum_i p(j|i)p_i$ and $q(i|j)=p(j|i)p_i/q_j$.
Under this map, $i$ is encoded into a quantum register, simply
containing the state $\ket{\ph_i}$ itself, and a classical
register, containing a classical mixture of $j$ values. Note that
this is a {\em single} signal encoding with {\em perfect} fidelity
since the state $\ket{\ph_i}$ may be regained perfectly from the
encoded version by simply discarding the classical register. Hence
our result characterizes optimal classical and quantum resources
in compression, in terms of very simple single-signal
perfect-fidelity encodings, despite the fact that compression is
defined asymptotically in terms of arbitrarily long signal strings
and fidelities merely {\em tending} to 1. This is a remarkable and
unexpected simplification --- even in classical information theory
it is by no means the rule that coding problems have solutions
that do {\em not} involve asymptotics (despite a few well known
examples such as Shannon's source and channel coding theorems
\cite{Shannon48}). The situation is even more tenuous in quantum
information theory, which seems to be plagued by further
non--additivity (or unresolved additivity questions) for some of
its basic quantities so that, at the present stage, many basic
constructions require a limit over optimization problems of
exponentially growing size.

Using our formula we will give a thorough discussion of further
properties of the trade--off curve including a generalization to
group covariant sources and to sources with infinitely many
(continuously parameterized) states. We show that our result also
leads to a number of corollaries characterizing the trade--off
between information gain and state disturbance for quantum sources
(yielding the results of~\cite{BarnumHJW01} on blind compression
as a corollary), and we indicate how our techniques for
characterizing $Q^*(R)$ provide a solution to the so-called remote
state preparation problem as well. Finally we develop a
probability--free version of our main result which may be
interpreted as an answer to the intuitive question: ``How many
classical bits does a qubit cost?'' This may also be interpreted
as a kind of dual to Holevo's theorem, insofar as the latter
characterizes the qubit cost of coding classical information into
qubits.

The presentation of these results is organized as follows. At the
top level, the paper is divided broadly into two parts. Part I,
comprising sections 2 through 8, sets up a precise formulation of
the basic definitions and the trade--off problem and gives the
proof of the main theorem characterizing $Q^*(R)$, as well as a
discussion of some of its important basic properties. Part II,
comprising sections 9 and 10, then goes on to provide some further
generalizations of the main result. In more detail, the contents
of the various sections are as follows.

In section \ref{sec:compression}, we will define the notions of
blind and visible compression, the essential difference being that
in the blind setting the encoder is given the actual quantum
states, while in the visible setting the encoder is given the
names of the quantum states as classical data. We then extend
these definitions to quantum--classical trade--off coding and
introduce the trade--off function $Q^*(R)$.

In section \ref{sec:lowerBound} we will prove a lower bound to the
trade--off curve in terms of the simple single--letter formula of
the ensemble decomposition construction paraphrased above. In
section \ref{sec:achieve} we will, in turn, show that the lower
bound is achievable so that the trade--off curve is identical to
the single--letter formula. This is our main result,
theorem~\ref{thm:central}.

In section \ref{sec:explore} we use our characterization of the
trade--off curve to evaluate $Q^*(R)$ numerically for a selection
of particular ensembles, chosen to illustrate various important
properties of the trade--off function. In section \ref{sec:avs} we
extend our results to a different asymptotic setting, known as the
arbitrarily varying source (AVS), in which there is no (or only
limited) knowledge of the prior probability distribution of the
states to be compressed. This provides a probability-free
generalization of our main result. In section
\ref{sec:infoDisturb} we show that our main result can be
reinterpreted to provide statements about the trade--off between
information gain and state disturbance for blind sources of
quantum states (in particular entailing a new proof of the main
result of \cite{BarnumHJW01}). Finally for part I, in section
\ref{sec:rsp} we indicate how our techniques -- developed to study
$Q^*(R)$ -- can also be used to characterize the trade--off curve
for the coding problem of remote state preparation posed in Refs.
\cite{lo:rsp} and \cite{rsp:prl}.
\par
Part II treats two significant further issues. In section
\ref{sec:symmetry} we show how to apply our results in the setting
of group covariant ensembles, which leads to considerable further
elegant simplifications. Section \ref{sec:infinitesource} is
devoted to the technicalities of generalizing our main result to
sources with infinitely many (continuously parameterized) states.
Finally, in an appendix, we collect proofs of various auxiliary
propositions that have been quoted in the body of the paper.\\[5mm]

{\Large\bf \begin{center} PART I:\\ CHARACTERIZING THE TRADE--OFF
CURVE
\end{center}}

\section{Blind and visible compression} \label{sec:compression}
We begin by introducing a number of definitions that are required
to give a precise statement of the variations of quantum source
coding that we will be considering in this paper. We will denote
an ensemble of quantum states $\ph_i$ with prior probabilities
$p_i$ as $\cE = \{ \ph_i, p_i \}$. In turn, we will write $S(\cE)
= S(\sum_i p_i \ph_i)$ for the von Neumann entropy of the average
state of the ensemble: $S(\rho)=-\Tr\rho\log\rho$. (Throughout
this paper $\log$ and $\exp$ will denote the logarithm and
exponential functions \emph{to base $2$}.) Starting from an
ensemble $\cE$, we can consider the quantum source producing
quantum states that are sequentially drawn independently from
$\cE$. Such a source corresponds to a sequence of ensembles
$\cE^{\ox n} = \{ \ph_I, p_I \}$, where
\begin{eqnarray}
I &:=& i_1 \ldots i_n \\
\ph_I &:=& \ph_{i_1} \ox \dots \ox \ph_{i_n} \\
p_I &:=& p_{i_1} \cdots p_{i_n}.
\end{eqnarray}
This sequence will be referred to as an independent identically
distributed (i.i.d.) source and the states of $\cE^{\ox n}$ are
called blocks of length $n$ from $\cE$. In this paper we will
focus on sources of pure quantum states $\ket{\ph_i}$, often
making use of the notation $\ph_i = \proj{\ph_i}$.  The measure
that we will use to determine whether two quantum states are close
is the fidelity $F$. For two mixed states $\r$ and $\o$, $F$ is
given by the formula
\begin{eqnarray}
F(\r,\o) := \left( \Tr \sqrt{\o^{1/2} \r \o^{1/2}} \right)^2.
\end{eqnarray}
(Note that some authors use the name ``fidelity'' to refer to the
square-root of this quantity.)  If $\o = \proj{\o}$ is a pure
state then the fidelity has a particularly simple form:
\begin{eqnarray}
F(\r,\o) = \bra{\o} \r \ket{\o} = \Tr(\r\o).
\end{eqnarray}
Finally, we will use the notation $\cH_d$ to denote the Hilbert
space of dimension $d$ and $\cB_d$ to denote the set of all mixed
states on $\cH_d$. Likewise, $\cH_d^{\ox n}$ will refer to the
$n$-fold tensor product of $\cH_d$ and, in a slight abuse of
notation, $\cB_d^{\ox n}$ will refer to the set of density
operators on $\cH_d^{\ox n}$. We are now ready to introduce the
definition of \emph{blind} quantum compression.
\begin{definition}
\label{Cdefn:quantumEncDec} A \emph{blind coding scheme}
for blocks of length $n$, to $R$ qubits per signal and fidelity
$1-\e$ comprises the following ingredients:
\begin{enumerate}
\item A completely positive, trace-preserving (CPTP) encoding map
$E_n : \cB_d^{\ox n} \rar \cB_2^{\ox nR}$.
\item A CPTP decoding map $D_n : \cB_2^{\ox nR} \rar \cB_d^{\ox n}$.
\end{enumerate}
such that average fidelity
\begin{equation}
\label{eq:fidelity}
\sum_I p_I \bra{\ph_I} D_n(E_n(\ph_I)) \ket{\ph_I} \geq 1-\e.
\end{equation}
We say that an i.i.d.~source $\cE$ can be blindly compressed to
$R$ qubits per signal if for all $\d,\e >0$ and sufficiently large
n there exists a blind coding scheme to $R+\d$ qubits per signal
with fidelity at least $1-\e$.
\end{definition}
The definition of visible compression is the same except that the
(CPTP) restrictions on the encoding map $E_n$ are relaxed; for
visible compression $E_n$ can be an arbitrary association of input
states to output states.  Equivalently, $E_n$ is a mapping from
the \emph{names} of the input states to output states.  Thus, we
write $E_n(I) \in \cB_2^{\ox nR}$.  Note that blind and visible
compression schemes differ only in the set of encoding maps that
are permitted. For blind (respectively visible) compression, the
input states are given as quantum (respectively classical)
information. In both cases the decoding must be CPTP. In this
language, the central result on the compression of quantum
information can be expressed as:
\begin{theorem}[Quantum source coding theorem~\cite{BFJS96,JozsaS94,schumacher}]
\label{thm:quantumsourcecoding}
A source $\cE$ of pure quantum states can be compressed to $\a$
qubits per signal if and only if $\a \geq S(\cE)$. The result
holds for both blind and visible compression.
\end{theorem}

It is interesting to study a refinement of quantum source coding
in which the states are coded into classical and quantum resources
which are quantified separately. Because of restrictions on the
manipulation of quantum states such as the no--cloning theorem
\cite{WoottersZ82}, blind compression is typically weaker than
visible.  In Refs. \cite{BarnumHJW01} and \cite{KoashiI}, for
example, it was shown that in blind compression it is typically
impossible to make use of classical storage.  The same is not true
in the visible setting, where it is possible to trade classical
storage for quantum. In this paper we study this trade--off for
{\em visible} compression but, before we begin, we need to recall
some basic definitions introduced in Ref. \cite{BarnumHJW01}.

Consider an encoding operation $E_n$ which maps a signal state
$\ket{\ph_I}$ into a joint state on a quantum register $B$ and a
classical register $C$. If $\{\ket{j}\}$ is the classical
orthonormal basis of $C$ then the most general classical state on
$C$ is a probability distribution over $j$ values, implying that
the most general form of the encoded state can be written as
\begin{eqnarray}
E_n(I) = \sum_j p(j|I) \o_{I,j}^B \ox \proj{j}^C.
\end{eqnarray}
The quantum and classical storage requirements (i.e. resources) of
the encoding map are simply the sizes of the registers $B$ and
$C$, respectively.
\begin{definition} \label{defn:quantumResource}
The \emph{quantum rate} of the encoding map $E_n$ is defined to be
$$\qsupp(E_n,\cE^{\ox n}) = \smfrac{1}{n} \log \dim \cH_B,$$
while the \emph{classical rate} of the encoding is defined to be
$$\csupp(E_n,\cE^{\ox n}) = \smfrac{1}{n} \log \dim \cH_C.$$
\end{definition}
With these definitions in place, we can make precise the notion of
compression with a quantum and a classical part.

\begin{definition}
\label{Cdefn:quantumResourceCompression}
A source $\cE$ can be compressed to $R$ classical bits per signal
plus $Q$ qubits per signal if for all $\e,\d > 0$ there exists an
$N>0$ such that for all $n>N$ there exists an encoding-decoding
scheme $(E_n,D_n)$ with fidelity $1-\e$ satisfying the
inequalities
\begin{eqnarray}
\csupp(E_n,\cE^{\ox n}) &\leq& R+\d, \\
\qsupp(E_n,\cE^{\ox n}) &\leq& Q+\d.
\end{eqnarray}
\end{definition}
The main result of this paper will be a complete characterization
of the curve describing the trade-off between $R$ and $Q$. As
mentioned above, for blind encodings there is usually no
trade--off to be made: generically, $Q \geq S(\cE)$, regardless of
the size of $R$. The reason is essentially that making effective
use of the classical register amounts to extracting classical
information from a quantum system in a reversible fashion, which
is impossible unless the quantum states of interest obey some
orthogonality condition. The more interesting case, therefore, is
to study the structure of the trade--off curve for visible
encodings.  As it turns out, our technique will yield the older
results for blind compression as a corollary.
\begin{definition} \label{Cdefn:tradeoffCurve}
For a given source $\cE = \{\ket{\ph_i},p_i\}$, define the
function $Q^*(R)$ to be the infimum over all values of $Q$ for
which the source can be visibly compressed to $R$ classical bits
per signal and $Q$ quantum bits per signal.
\end{definition}
Some properties of the curve $Q^*(R)$ are immediate. For example,
the endpoints of the curve are easily found.  If $R=0$ then the
compression must be fully quantum mechanical and the quantum
source coding theorem~\ref{thm:quantumsourcecoding} applies:
$Q^*(0)=S(\cE)$.  More generally, the theorem implies that
$Q^*(R)+R \geq S(\cE)$ for all $R$. Similarly, for $R=H(p)$ we
have $Q^*(R)=0$, by Shannon's classical source coding theorem.
Moreover, for intermediate values of $R$, the curve is necessarily
convex because one method of compressing with classical rate $\l_1
R_1 + \l_2 R_2$ is simply to timeshare between the optimal
protocols for $R_1$ and $R_2$ individually, resulting in quantum
rate of $\l_1 Q^*(R_1) + \l_2 Q^*(R_2)$.
\par\medskip
\begin{example}\label{BB84example} \exampleTitle{Parameterized BB84 ensemble}
Let us consider in more detail the  example of a parameterized
version of the BB84 ensemble in order to see what sorts of
protocols are possible beyond simple time-sharing.  For $0 < \t
\leq \p/4$, let $\cE_{BB}(\t)$ be the ensemble consisting of the
states
\begin{eqnarray}
\ket{\ph_1} &=& \ket{0} \\
\ket{\ph_2} &=& \cos\t \ket{0} + \sin\t \ket{1} \\
\ket{\ph_3} &=& \ket{1} \\
\ket{\ph_4} &=& -\sin\t \ket{0} + \cos\t \ket{1},
\end{eqnarray}
as illustrated in figure \ref{fig:BB84:pic}, each occurring with
probability $p_i= 1/4$. We then have $S(\cE)= 1$ and $H(p)=2$.
From the argument above, we therefore already know two points on
the $(R,Q^*(R))$ curve, namely $(0,1)$ and $(2,0)$.  To get a
better upper bound than the straight line joining these two
points, suppose we were to partition the four states into two
subsets, $\cX_1 = \{ \ket{\ph_1}, \ket{\ph_2} \}$ and $\cX_2 = \{
\ket{\ph_3}, \ket{\ph_4} \}$.  For a given input string $I = i_1
i_2 \dots i_n$, the classical register could be used to encode,
for each $k$, whether $\ket{\ph_{i_k}} \in \cX_1$ or
$\ket{\ph_{i_k}} \in \cX_2$.  The classical rate required to do so
would be $1$ classical bit per signal.  Independent of the value
of the classical register, the quantum resource required to
compress the subensembles is then just the quantum resource
required to compress a pair of equiprobable quantum states
subtended by the angle $\t$. Therefore,
\begin{eqnarray}
Q^*(1)
&\leq& S\left( \smfrac{1}{2} \proj{\ph_1} + \smfrac{1}{2} \proj{\ph_2} \right)
= H_2\left(\smfrac{1}{2}(1 + \cos \t) \right).
\end{eqnarray}
By time-sharing between the point corresponding to this protocol
and the two endpoints of the curve that we already calculated, we
get a piecewise linear upper bound on $Q^*$.  As we will see
later, however, the true curve is strictly below this upper bound.
(The impatient reader is allowed to peek at figure
\ref{fig:BB84:graph} in section \ref{sec:explore}.)
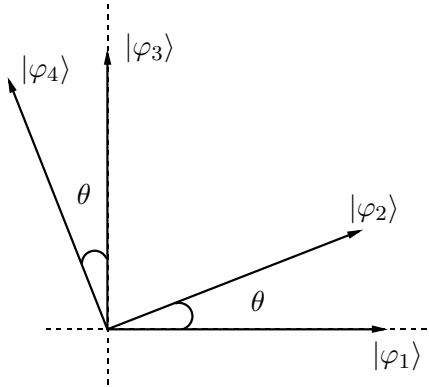
\begin{figure}
\begin{center}
\setlength{\unitlength}{0.00043333in}
{\renewcommand{\dashlinestretch}{30}
\begin{picture}(5160,4689)(0,-10)
\thicklines \put(1090.500,1489.500){\arc{315.357}{3.1892}{5.8408}}
\put(2123.806,841.597){\arc{329.495}{4.4709}{7.4837}}
\thinlines
\thicklines \path(1248,687)(4623,687)
\blacken\path(4503.000,657.000)(4623.000,687.000)(4503.000,717.000)(4503.000,657.000)
\path(1248,687)(1248,4062)
\blacken\path(1278.000,3942.000)(1248.000,4062.000)(1218.000,3942.000)(1278.000,3942.000)
\path(1248,687)(4323,1887)
\blacken\path(4222.117,1815.428)(4323.000,1887.000)(4200.304,1871.322)(4222.117,1815.428)
\path(1248,687)(44,3722)
\blacken\path(115.572,3621.117)(44.000,3722.000)(59.678,3599.304)(115.572,3621.117)
\thinlines \dashline{60.000}(1248,4662)(1248,12)
\dashline{60.000}(498,687)(5148,687) \put(873,2217){$\theta$}
\put(3003,912){$\theta$} \put(183,3777){$\ket{\ph_4}$}
\put(1458,4047){$\ket{\ph_3}$} \put(4203,2077){$\ket{\ph_2}$}
\put(4488,267){$\ket{\ph_1}$}
\end{picture}
}
\end{center}
\caption{Parameterized BB84 ensemble $\cE_{BB}(\t)$.}
\label{fig:BB84:pic}
\end{figure}

\end{example}
With this example in mind, let us move on to our analysis of the
general case.

\section{Single--letter lower bound on $Q^*(R)$} \label{sec:lowerBound}
In this section we will prove a lower bound on the
quantum--classical trade--off curve by reducing the asymptotic
problem to a single--copy problem. Because compression is only
possible asymptotically, however, we need to shift the emphasis
away from the quantum and classical resources towards quantum and
classical mutual information quantities.  In the next section we
will then prove that nothing was lost by making this shift -- we
will show that the resulting lower bound to $Q^*(R)$ is actually
achievable.

\subsection{Mutual information and additivity}\label{subsec:mutualInfo}
The information quantities in question will be the mutual
informations between the name of the state being compressed and
the quantum and classical registers containing the output of the
encoding map $E_n$.  Thus, we define the state
\begin{equation} \label{eqn:encodedState}
\r^{ABC} :=
\sum_{I,j} p_I \proj{I}^A \ox p(j|I) \o_{I,j}^B \ox \proj{j}^C.
\end{equation}
The names $I$ are stored in orthogonal states on system $A$ while
the quantum and classical encoding registers are labelled $B$ and
$C$, respectively.  We can then make the following definitions:
\begin{eqnarray}
S(A:C) &:=& S(A) + S(C) - S(AC) \\ S(A:B|C) &:=&
S(AC)+S(BC)-S(ABC)-S(C),
\end{eqnarray}
where, for any subsystem $X$, $S(X)$ denotes the von Neumann
entropy of the reduced state of $X$. Note that $S(A:C)$ is just
the classical mutual information $H(I:j)$ between $I$ and $j$. To
interpret $S(A:B|C)$, observe that for a given classical output
$j$, we can write down a conditional ensemble
\begin{eqnarray}
\cE_j = \{ \o_{I,j}, q(I|j) \},
\end{eqnarray}
where $q(I|j)$ is calculated using Bayes' rule to be $q(I|j) =
p(j|I)p_I/q_j$, with $q_j = \sum_I p(j|I) p_I$.  The conditional
quantum mutual information $S(A:B|C)$ is just the average Holevo
information $\chi$ of the conditional ensembles $\cE_j$:
\begin{equation}
S(A:B|C) = \sum_j q_j \c(\cE_j),
\end{equation}
where $\c$ is defined, for an ensemble $\cE=\{\rho_k,p_k\}$, as
\cite{Holevo73}
\begin{equation}
\c(\cE):=S\left( \sum_k p_k\rho_k \right)-\sum_k p_k S(\rho_k).
\end{equation}
Because $\cE_j$ is an ensemble supported on system $B$, $\c(\cE_j)
\leq n \qsupp$, which implies that
\begin{equation} \label{eqn:basicBound}
n \, \qsupp \geq S(A:B|C).
\end{equation}
Therefore, roughly speaking, we will derive a lower bound on
$Q^*(R)$ by minimizing $S(A:B|C)$ subject to the constraint
$S(A:C) \leq nR$ and developing further properties of that
minimum. To that end, define $T_\e(\cE^{\ox n},nR)$ to be the set
of all encoding maps $E$ for which $S(A:C)\leq nR$ and there
exists a decoding map $D$ satisfying
\begin{equation}
\sum_I p(I) F(\r_I,(D \circ E)\r_I) \geq 1- \e.
\end{equation}
Next define $M_\e(\cE^{\ox n},nR)$ to be the infimum of $S(A:B|C)$
over all $E \in T_\e(\cE^{\ox n},nR)$. We begin by noting the
following basic properties of $M_\e(\cE,R)$.
\begin{lemma}
\label{lemma:Mproperties}
$M_\e(\cE,R)$ is a monotonically decreasing function of $R$.
Moreoever, it is jointly convex in $\e$ and $R$, in the sense
that, for any set of $\e_k > 0$ and $R_k \geq 0$ as well as
probabilities $\sum_k \l_k = 1$,
\begin{eqnarray}
M_\e(\cE,R) \leq \sum_k \l_k M_{\e_k}(\cE,R_k),
\end{eqnarray}
where $\e = \sum_k \l_k \e_k$ and $R = \sum_k \l_k R_k$.
\end{lemma}
\begin{proof}
Monotonicity follows immediately from the definitions. If $R_1
\leq R_2$ and $S(A:C) \leq R_1$ then $S(A:C) \leq R_2$. Thus
the set $T_\e(\cE,R_1)$ is contained in $T_\e(\cE,R_2)$ and
$M_\e(\cE,R_1) \geq M_\e(\cE,R_2)$.

To prove joint convexity, let $\e_k$, $R_k$ and $\l_k$ be as in
the statement of the lemma and assume that $E_k \in
T_{\e_k}(\cE,R_k)$. Furthermore, suppose that the encoding maps
$E_k$ each have separate, distinguishable classical registers
$C_k$. We construct an encoding map with information rate $R \leq
\sum_k \l_k R_k$ and fidelity $\e \leq \sum_k \l_k \e_k$ by
applying the map $E_k$ with probability $\l_k$. The first
inequality follows from the fact that the registers $C_k$ are
separate:
\begin{equation}
S(A:C) = \sum_k \l_k S(A:C_k) \leq R.
\end{equation}
The decoding map for the new encoding consists of first
determining which classical register $C_k$ was used and then
applying the decoding map corresponding to $E_k$.  The output of
the encoding--decoding scheme will, therefore, be the average of
the outputs of the individual schemes, yielding $1 - \e \geq
\sum_k \l_k ( 1 - \e_k )$ by the concavity of the fidelity.
Finally, if we define $S_k(A:B|C)$ to be the conditional quantum
mutual information for the encoding map $E_k$ then we can
calculate the value for the new scheme,
\begin{eqnarray}
S(A:B|C) = \sum_k \l_k S_k(A:B|C).
\end{eqnarray}
Since $M_\e(\cE,R) \leq S(A:B|C)$ by definition and this
inequality must hold for all encoding maps $E_k$, we can conclude
that $M_\e(\cE,R) \leq \sum_k \l_k M_\e(\cE,R_k)$.
\end{proof}
The particular usefulness of the $M_\e$ function derives from an
additivity property with respect to the input ensemble given in
the next lemma, a property that can be converted into a
single--letter lower bound on $Q^*(R)$.
\begin{lemma} \label{lemma:additivity}
For any ensemble $\cE$, numbers $R, \e \geq 0$ and non-negative
integer $n$,
\begin{equation}
M_\e(\cE^{\ox n},nR) \geq n M_\e(\cE,R).
\end{equation}
\end{lemma}
\begin{proof}
To begin, recall that $I=i_1 i_2 \dots i_n$ and decompose $A$ into
$A_1 A_2 \dots A_n$, with $\ket{i_k}$ stored on $A_k$. We will
frequently make use of the notation $A_{<k} = A_1 A_2 \dots
A_{k-1}$ and the analogous $I_{<k} = i_1 i_2 \dots i_{k-1}$, as
well the similar $A_{>k}$ and $I_{>k}$. For a fixed $E \in
T_\e(\cE^{\ox n},nR)$, the chain rule for mutual information (cf.
appendix C of Ref. \cite{BarnumHJW01}) implies that
\begin{eqnarray}
S(A:B|C) &=& \sum_{k=1}^n S(A_k:B|C,A_{<k}). \label{eqn:breakup}
\end{eqnarray}
The bulk of the proof will consist of definitions for the purpose
of interpreting the individual summands in the chain rule in terms
of single--copy encoding maps. Consider one such term,
$S(A_k:B|C,A_{<k})$, which we can express as
\begin{eqnarray}
S(A_k:B|C,A_{<k}) &=& \sum_{I_{<k},j} p(I_{<k},j) \c(
\cE_{I_{<k},j} ),
\end{eqnarray}
where $\cE_{I_{<k},j}$ is the ensemble of states
\begin{equation}
\cE_{I_{<k},j} = \left\{ \sum_{I_{>k}} p(I_{>k}) \o_{I,j}, q_{I_{<k}}(i_k|j) \right\},
\end{equation}
with
\begin{equation}
q_{I_{<k}}(i_k|j) = \frac{\sum_{I_{>k}}p(i_k)p(I_{>k})p(j|I)}{\sum_{I_{\geq k}}p(I_{\geq k})p(j|I)}.
\end{equation}
Now define the encoding map $E_{I_{<k}}$ on the ensemble $\cE$ to be
\begin{equation}
E_{I_{<k}}(i_k):= \sum_{I_{>k}} p(I_{>k}) E(I)
                = \sum_{I_{>k}}\sum_j p(I_{>k})p(j|I)\omega_{I,j} \ox \proj{j}.
\end{equation}
The output of
$E_{I_{<k}}$ on the quantum register is described by the set of ensembles
$\cE_{I_{<k},j}$. Next, define the decoding map $D_k = \Tr_{\neq k} \circ D$
and the fidelity
\begin{equation}
F_{I_{<k}} := 1 - \e_{I_{<k}} := \sum_{i_k}
    p(i_k) F\bigl(\r_{i_k},(D_k \circ E_{I_{<k}})(i_k)\bigr).
\end{equation}
We can then calculate that
\begin{eqnarray}
\sum_{I_{<k}} p(I_{<k}) F_{I_{<k}}
&=& \sum_{I_{<k}} p(I_{<k}) \sum_{i_k} p(i_k) F \left( \r_{i_k},
        (D_k \circ E_{I_{<k}})(i_k) \right)  \\
&=& \sum_{I_{\leq k}} p(I_{\leq k})
    F\!\left( \r_{i_k},
        \Tr_{\neq k}D\!\left( \sum_{I_{>k}} p(I_{>k}) E(I) \right) \right)
    \nonumber \\
&=& \sum_{I_{\leq k}} p(I_{\leq k})
    F\!\left( \sum_{I_{>k}} p(I_{>k}) \r_{i_k},
        \sum_{I_{>k}} p(I_{>k}) (\Tr_{\neq k}\circ D\circ E)(I) )\right)
    \nonumber \\
&\geq& \sum_I p(I) F\bigl( \Tr_{\neq k}\r_I, (\Tr_{\neq k}\circ D \circ E)(I) \bigr) \nonumber \\
&\geq& \sum_I p(I) F\bigl( \r_I, (D \circ E)(I) \bigr) \nonumber \\
\label{eq:crucial:lower}
&\geq& 1 - \e. \nonumber
\end{eqnarray}
The first three lines are by definition and using linearity to
shuffle the terms.  The first inequality comes from the joint
concavity of the fidelity, the second from its monotonicity under
partial trace, and the last from the fidelity condition on $D
\circ E$.

Therefore, if we write $j(E_{I_{<k}})$ for the random variable
representing the classical output of the encoding map $E_{I_{<k}}$
and $R_{I_{<k}}$ for the corresponding mutual information then
$E_{I_{<k}} \in T_{\e_{I_{<k}}}(\cE,R_{I_{<k}})$.  Defining $R_k
:= \sum_{I_{<k}} p(I_{<k}) R_{I_{<k}}$ for the average classical
information and applying the joint convexity of $M$ then finally
yields
\begin{eqnarray} \label{eqn:singleTerm}
S(A_k:B|C,A_{<k}) \geq M_\e(\cE,R_k).
\end{eqnarray}
A simple calculation allows us to bound the $R_k$ from above,
however:
\begin{eqnarray}
\sum_k R_k
&=& \sum_k \sum_{I_{<k}} p(I_{<k}) H\bigl(i_k:j(E_{I_{<k}})\bigr) \\
&=& \sum_k S(A_k:C|A_{<k}) \\
&=& S(A:C) \leq nR. \label{eqn:rateConstraint}
\end{eqnarray}
Combining Eqs. (\ref{eqn:singleTerm}) and
(\ref{eqn:rateConstraint}) with the chain rule, and applying the
convexity of $M$ one more time gives the simple inequality
\begin{equation}
\label{eq:finishoff}
S(A:B|C) \geq \sum_k M_\e(\cE,R_k) \geq n M_\e(\cE,R).
\end{equation}
Since this lower bound must hold for all encoding maps in
$T_\e(\cE^{\ox n},R)$, that concludes the proof of the lemma.
\end{proof}

\subsection{Perfect encodings and their properties}
\label{subsec:perfectEncodings}
Within the set $T_0(\cE,R)$ of encoding maps with {\em perfect}
fidelity decodings there is a particularly simple subset, in terms
of which we will phrase our final bound on $Q^*(R)$.  Let
$T(\cE,R) \subset T_0(\cE,R)$ be the set of all encoding maps $E$
of the form
\begin{eqnarray}\label{specenc}
E(i) = \proj{\ph_i}^B \ox \sum_j p(j|i) \proj{j}^C.
\end{eqnarray}
In other words, $T(\cE,R)$ consists of the encoding maps in which
a perfect copy of the state to be compressed is placed in register
$B$. The decoding map is simply to trace over the register $C$.
While such encodings, which simply reproduce the input, are
obviously useless for compression, they turn out to be quite
sufficient for minimizing $S(A:B|C)$.  Indeed, let us define
\begin{eqnarray}
M(\cE,R) &=& \inf \{ S(A:B|C) : E \in T(\cE,R) \} \\
         &=& \inf_{p(\cdot|\cdot)} \{ S(A:B|C) : S(A:C)\leq R
         \}.\label{defMER}
\end{eqnarray}
By construction, this optimization is no longer over general CPTP
maps but only over different possible conditional probability
distributions on register $C$.
\par\medskip
Let us collect a few properties of $M$ for later use: First of
all, $M$ inherits the convexity of $M_\e$ in the variable $R$.
Also, it clearly is nonincreasing, and $M(\cE,0)=S(\cE)$ is
immediate from the definition. Furthermore, for any choice of
$p(\cdot|\cdot)$, we have
\begin{equation}
S(A:C)+S(A:B|C)
           =    S(A:BC)
           \geq S(A:B) =S(\cE),
\end{equation}
from which we conclude that $R+M(\cE,R) \geq S(\cE)$. This,
together with the convexity, implies continuity in $R$, and the
estimates
\begin{equation}
\label{eq:M:continuity}
M(\cE,R) \geq M(\cE,R+\delta) \geq M(\cE,R)-\delta.
\end{equation}
In what follows, it will also frequently be helpful to use the
following fact:
\begin{proposition}\label{prop:useEquality}
\begin{eqnarray}
M(\cE,R) &=& \inf_{p(\cdot|\cdot)} \{ S(A:B|C) : S(A:C) = R \},
\end{eqnarray}
with an equality condition in the infimum (rather than the inequality
of Eq. (\ref{defMER})).
\end{proposition}
The proof is given in appendix \ref{pfprop:useEquality}.

In principle one might envisage a limit with larger and larger
classical register $C$. This would constitute a serious obstacle
to calculating $M(\cE,R)$ and carrying through our larger program
of evaluating $Q^*(R)$.  Fortunately, the next proposition ensures
that the range of $j$'s we need to consider in the definition of
$M(\cE,R)$ is bounded universally. Since the mutual informations
involved are continuous, the infimum in the definition of
$M(\cE,R)$ can be replaced by a minimum.

\begin{proposition} \label{prop3.two} In the definition of $M
(\cE,R)$ given in Eq. (\ref{defMER}), it suffices to consider
encodings of the form Eq. (\ref{specenc}) with at most $(m+1)$\, $j$
values, where $m$ is the number of states in $\cE$.
\end{proposition}
The proof is given in appendix \ref{pfprop3.two}.

\subsection{Completing the lower bound}\label{subsec:completingLB}
Returning to the main argument, we are now prepared to relate
$M(\cE,R)$ to the trade--off curve:
\begin{theorem}
\label{thm:lowerbound}
If a source $\cE$ can be visibly compressed to $Q$ qubits per
signal and $R$ classical bits per signal then $Q \geq M(\cE,R)$.
Equivalently, $Q^*(R) \geq M(\cE,R)$.
\end{theorem}
\begin{proof}
By the definition of compression and the previous lemma, we note
that, for all $\e,\d > 0$, the inequality $Q^*(R) \geq
M_{\e}(\cE,R+\delta)$ must hold. We will give a proof that $M_\e$
is continuous at $\e=0$, from which the stronger lower bound in
terms of $M(\cE,R)$ will follow.

So, fix $\e,\d$ for now and suppose that $E \in T_\e(\cE,R+\d)$.
Let $D$ be the decoding map associated to $E$.  As usual,
\begin{equation}
E(i) = \sum_j \o_{i,j}^B \ox p(j|i) \proj{j}^C.
\end{equation}
For a given $j$ value, the decoding map will produce the ensemble
of states $\{\s_{i,j},p(i|j)\}$ where $\s_{i,j} =
D(\o_{i,j}^B\ox\proj{j}^B)$. Therefore, applying Markov's
inequality (cf. lemma 6.3 of Ref. \cite{BarnumHJW01})  and the
fidelity condition in the definition of $T_\e(\cE,R)$, the
probability weight of the $j$'s with
\begin{equation}
\sum_i q(i|j) F(\ph_i,\s_{i,j}) \geq 1-\sqrt{\e}
\end{equation}
is at least $1-\sqrt{\e}$.  In other words, for these good $j$
values, the output of the decoding map is close to $\cE_j$.
Therefore, for these same good $j$ values, by the monotonicity and
continuity of $\c$, we must have
\begin{equation}
\c(\cE_j) \geq S\left( \sum_i q(i|j) \proj{\ph_i} \right) - f(\e),
\end{equation}
where we may choose $f(\e)=4(\sqrt[4]{\e}\log d -\sqrt[4]{\e} \log
(2\sqrt[4]{\e}))$ (as shown in appendix A of Ref.
\cite{BarnumHJW01}). Consequently,
\begin{equation}
S(A:B|C) = \sum_j q_j \c(\cE_j)
         \geq \sum_j q_j S\left( \sum_i q(i|j) \proj{\ph_i} \right) -
         f(\e).
\end{equation}
Since $f(\e)\rightarrow 0$ as $\e \rightarrow 0$ we conclude that
$\lim_{\e\dn 0} M_\e(\cE,R+\d) = M_0(\cE,R+\d)$ and, moreover, in the
limit $\e \rar 0$ it suffices to consider encoding maps of the type
\begin{equation}\label{encspecial}
E(i) = \proj{\ph_i}^B \ox \sum_j p(j|i) \proj{j}^C.
\end{equation}
Thus we obtain $Q^*(R) \geq M(\cE,R+\d)$, for all $\d>0$, which,
by Eq.~(\ref{eq:M:continuity}) above yields our claim.
\end{proof}
\par
\medskip
\begin{remark}
The estimate $f(\e)$ above may also be derived using Fannes'
inequality~\cite{Fannes}, which states that
 for density operators $\rho$ and $\sigma$ on a $d$--dimensional
space,
\begin{equation}
\label{eq:fannes} \|\rho-\sigma\|_1\leq\e \Longrightarrow
|S(\rho)-S(\sigma)|\leq d\eta(\e/d).
\end{equation}
where
\begin{equation}\label{etaeqn}
  \eta(x)=\begin{cases}
            -x\log x    & \text{ for }x\leq\frac{1}{4}, \\
            \frac{1}{2} & \text{ for }x >  \frac{1}{4}.
          \end{cases}
\end{equation}
We will use this inequality again later.
\end{remark}

\subsection{On alternative definitions}\label{subsec:alternativeDefs}
Inspecting the proofs of lemma \ref{lemma:additivity} and theorem
\ref{thm:lowerbound} reveals that we do not actually need the
block--based fidelity condition
\begin{equation}
\avg{F} :=\sum_I p_I F\bigl( \r_I,(D \circ E)(I)\bigr) \geq 1-\e,
\end{equation}
of Eq.~(\ref{eq:fidelity}) but only the weaker mean letterwise fidelity
\begin{equation}
\label{eq:letterfidelity}
\avg{\overline{F}} := \sum_I p_I \overline{F}_I \geq 1- \e,
\end{equation}
where
\begin{equation}
\overline{F}_I := \frac{1}{n}
                 \left[ \sum_{k=1}^n F\bigl( \r_{i_k},(\Tr_{\neq k}\circ D \circ E)(I) \bigr)
                        \right].
\end{equation}
By the monotonicity of the fidelity under partial traces, the latter is
directly implied by the former.

The lower bound Eq.~(\ref{eq:crucial:lower}) is then replaced by
$1-\e_k$, with $\frac{1}{n}\sum_k \e_k=\e$, and we conclude, instead
of Eq.~(\ref{eqn:singleTerm}), that
\begin{equation}
S(A_k:B|C,A_{<k}) \geq M_{\e_k}(\cE,R_k).
\end{equation}
The remaining argument is only altered at Eq.~(\ref{eq:finishoff}):
\begin{equation}
S(A:B|C) \geq \sum_{k=1}^n M_{\e_k}(\cE,R_k) \geq nM_{\e}(\cE,R),
\end{equation}
using joint convexity once more.
\par
Hence, we could define the function $\lbar{M}_\e(\cE,R)$ in a
fashion analogous to $M_\e(\cE,R)$ but using the fidelity function
$\lbar{F}$ instead of $F$ and lemma \ref{lemma:additivity} would
continue to hold for the new function. In fact,
$\lbar{M}_\e(\cE,R)$ will be strictly additive, in the sense that
\begin{eqnarray}
\lbar{M}_\e(\cE^{\ox n},nR) = n\lbar{M}_\e(\cE,R),
\end{eqnarray}
because any single letter encoding with fidelity $1-\e$ repeated
$n$ times gives rise to an $n$-block coding with mean letterwise
fidelity $1-\e$.

We also note at this stage that we could have opted for a slightly
more sophisticated definition of the quantum resource of the
encoding.  In particular, if we introduce $\qsupp_j
= \smfrac{1}{n} \log \Rank \cE_j$ as the minimal number of qubits
per signal required to support the conditional ensemble $\cE_j$
then we could have defined the quantum rate of the encoding map as
\begin{equation}
\lbar{\qsupp} = \sum_j q_j \; \qsupp_j.
\end{equation}
In this picture, the quantum resource would be the average over
classical $j$ values of the minimal number of qubits per signal
required to support the quantum portion of the encoded state
$E_n(I)$.  Such a definition, by treating the classical and quantum
storage requirements differently, allows the possibility of
variable--length quantum encodings, where the length is a function of
the classical message $j$.  Such encodings could potentially be more
powerful than the encodings with fixed--sized quantum supports used
to define the original $\qsupp$.  However, because $\qsupp_j \geq
\c(\cE_j)$, the analog of Eq.~(\ref{eqn:basicBound}) continues to
hold.  (For a more detailed investigation of the properties of
such variable--length quantum memories, see \cite{Kuperberg}.)
More precisely,
\begin{eqnarray}
n \, \overline{\qsupp} \geq S(A:B|C).
\end{eqnarray}
Therefore, the lower bound of theorem \ref{thm:lowerbound} on the
trade--off curve $Q^*(R)$ would apply equally well if we had defined
$Q^*(R)$ using $\overline{\qsupp}$ instead of $\qsupp$.

Thus, while replacing either $F$ by $\lbar{F}$ or $\qsupp$ by
$\overline{\qsupp}$ in the definition of compression could
potentially have reduced the resource requirements, we find that
our lower bounds would apply to the modified definitions.  Since
we will see later in the paper that the lower bounds are
achievable using the original, restrictive formulation of
compression, we can conclude that no advantage can be gained by
relaxing the definitions to use $\lbar{F}$ and
$\overline{\qsupp}$.

\section{Achieving the lower bound $M(\cE,R)$} \label{sec:achieve}
Recall that the trade--off function $Q^* (R)$ gives the minimal
quantum resource $Q^*$ qubits per letter that is sufficient to
encode arbitrarily long strings with arbitrarily high fidelity
$1-\e$ for any $\e >0$, given a classical resource of $R$ bits per
letter. On the other hand the lower bound $M(\cE,R)$ is defined as
the minimal quantum resource for a particular kind of {\em single}
letter {\em perfect} fidelity (i.e. $\e =0$) encoding given in Eq.
(\ref{encspecial}), subject to the constraint that the classical
{\em mutual information} $S(A:C)$ between $i$ and $j$ is $R$.
Hence in the latter case, the classical resource will generally
exceed $R$ bits per letter. Thus by implementing the simple
encodings of Eq. (\ref{encspecial}) we can attain $M(\cE,R)$ as
the quantum resource but not generally with a classical resource
bounded by $R$. We now argue that nevertheless, the classical
resource can be reduced to $R$ while retaining the quantum
resource at $M(\cE,R)$ i.e. that the lower bound $M(\cE,R)$ to
$Q^*(R)$ is attainable, so we must then have $Q^*(R)=M(\cE,R)$.

Our strategy intuitively is the following. We think of the
conditional distribution $p(j|i)$ with mutual information $S(A:C)$ in
Eq. (\ref{encspecial})  as a noisy channel from $i$ to $j$. Then the
reverse Shannon theorem \cite{bennett:et:al} states that this noisy
channel can be simulated with a noiseless channel of capacity
$S(A:C)$ if the receiver and sender have shared randomness i.e. in
the presence of shared randomness, the classical resource can be
reduced to $R=S(A:C)$ bits per letter. Finally we show that only
$O(\log n)$ bits of shared randomness suffice to provide a high
fidelity encoding-decoding scheme for blocks of length $n$. Hence
this amount of shared randomness can be included in the classical
resource of the encoding with asymptotically vanishing cost per
letter.

To make the above intuitions mathematically rigorous, we begin by
recalling some basic facts from the theory of typical
sequences~\cite{csiszar:koerner,wolfowitz} and typical
subspaces~\cite{schumacher,winter:diss} in the following two
subsections.

\subsection{Typical sequences}
\label{subsec:typ:seq}
For a sequence $I=i_1\ldots i_n\in\cI^n$ define the \emph{type} $P_I$
of $I$ as its empirical distribution of letters, i.e.
\begin{equation}
P_I(i):=\frac{1}{n}N(i|I)
      :=\frac{1}{n}|\{k|i_k=i\}|.
\end{equation}
The number of types of sequences is polynomial in $n$: it is
$\binom{n+|\cI|-1}{|\cI|-1} \leq (n+1)^{|\cI|}$.
\par
The \emph{type class} $\cT_P$ of $P$ is the set of all sequences
with type $P$:
\begin{equation}
\cT_P:=\{ I\in\cI^n | P_I=P \}.
\end{equation}
Consider now any probability distribution $P$ on $\cI$, and let $\delta>0$.
Then the set of \emph{typical sequences} (with respect to the distribution $P$
and $\delta$) is
\begin{equation}
\cT_{P,\delta}:=\bigl\{ I\in\cI :
                  \forall i\ |P_I(i)-P(i)| \leq \delta/\sqrt{n} \bigr\}.
\end{equation}
Note that this set is a union of certain type classes.
\par
The following are standard facts~\cite{csiszar:koerner,wolfowitz}:
\begin{equation}
P^{\otimes n}(\cT_{P,\delta}) \geq 1-\frac{1}{\delta^2}.
\end{equation}
\begin{eqnarray}
(n+1)^{-|\cI|}\exp\bigl( n(H(P)) \bigr) &\leq & |\cT_P|,          \\
              \exp\bigl( n(H(P)) \bigr) &\geq & |\cT_P|,          \\
(n+1)^{-|\cI|}\exp\bigl( n(H(P)-|\cI|\eta(\d/\sqrt{n})) \bigr)
                                        &\leq & |\cT_{P,\d}|,     \\
 (n+1)^{|\cI|}\exp\bigl( n(H(P)+|\cI|\eta(\d/\sqrt{n})) \bigr)
                                        &\geq & |\cT_{P,\d}|.
\end{eqnarray}
Note that the latter two follow from the former two by the following
well--known explicit estimate on the difference of two
entropies~\cite{csiszar:koerner} (this being a classical case of the
Fannes inequality, Eq.~(\ref{eq:fannes})): if $P$ and $Q$ are
probability distributions on a set of $k$ elements then
\begin{equation}
\|P-Q\|_1\leq \e \Longrightarrow |H(P)-H(Q)| \leq
k\eta\left(\frac{\e}{k}\right)
\end{equation}
where the function $\eta$ is given in Eq. (\ref{etaeqn}).
\par\medskip
For sequences $I\in\cI^n$, $J\in\cJ^n$, the \emph{conditional type} $W_{J|I}$ of
$J$ (conditional on $I$) is defined as the stochastic matrix given by
\begin{equation}
\forall ij\quad P_I(i)W_{J|I}(j|i) = P_{IJ}(ij),
\end{equation}
where $P_{IJ}$ is the joint type of $IJ=(i_1j_1,\dots,i_nj_n)$. It is
undetermined if $P_I(i)=0$.
\par
The \emph{conditional type class} of $W$ given $I$ is defined as
\begin{equation}
\cT_W(I) := \{ J : W_{J|I}=W \}
          = \{ J : \forall ij\ P_{IJ}(ij)=P_I(i)W(j|i) \}.
\end{equation}
Let $W$ be now an arbitrary stochastic matrix and $\d>0$.
The \emph{set of conditionally typical sequences} of $W$ given $I$
is defined as
\begin{equation}
\label{eq:weak:law}
\cT_{W,\d}(I) := \bigl\{ J : \forall ij\
            |W_{J|I}(j|i)-W(j|i)| \leq \delta/\sqrt{N(i|I)} \bigr\}.
\end{equation}
Again, there are a couple of standard facts:
\begin{equation}
W_I\bigl(\cT_{W,\d}(I)\bigr) \geq 1-\frac{|\cI|}{\d^2},
\end{equation}
for the product distribution $W_I=W_{i_1}\ox\cdots\ox W_{i_n}$, and
\begin{eqnarray}
(n+1)^{-|\cI||\cJ|}\exp\bigl( nH(W|P_I) \bigr) &\leq & |\cT_W(I)|,           \\
                   \exp\bigl( nH(W|P_I) \bigr) &\geq & |\cT_W(I)|,           \\
(n+1)^{-|\cI||\cJ|}\exp\bigl( n(H(W|P_I)-|\cI||\cJ|\eta(\d|\cI|/\sqrt{n})) \bigr)
                                               &\leq & |\cT_{W,\d}(I)|,       \\
\label{eq:conditional:typical:upper}
 (n+1)^{|\cI||\cJ|}\exp\bigl( n(H(W|P_I)+|\cI||\cJ|\eta(\d|\cI|/\sqrt{n})) \bigr)
                                               &\geq & |\cT_{W,\d}(I)|,
\end{eqnarray}
where $H(W|P_I)$ is just the conditional Shannon entropy
$\sum_i P_I(i) H\bigl(W(\cdot|i)\bigr)$.

\subsection{Typical subspaces}
\label{subsec:typ:space}
The concepts in the previous subsection translate straightforwardly
to their Hilbert space versions via the following recipe:
\par
For a state $\rho$ choose a diagonalization $\rho=\sum_{i\in\cI}
r_i\proj{e_i}$, with eigenvectors $\ket{e_i}$ and eigenvalues $r_i$,
which define a probability distribution on $\cI$. Then we have a
diagonalization of $\rho^{\otimes n}$:
\begin{equation}
\rho^{\otimes n}=\sum_{I\in\cI} r_I\proj{e_I},
\end{equation}
with
\begin{align}
  \ket{e_I} &= \ket{e_{i_1}}\otimes\cdots\otimes\ket{e_{i_n}}, \\
  r_I       &= r_{i_1}\cdots r_{i_n}.
\end{align}
Now for any subset $\cA\subset\cI^n$ we can define the subspace spanned
by the vectors $\{\ket{e_I} : I\in\cA \}$, which is most conveniently
described by the subspace projector
\begin{equation}
\Pi_{\cA} := \sum_{I\in\cA} \proj{e_I}.
\end{equation}
In this way we can define, for any distribution $P$ on $\cI$,
\begin{equation}
\Pi_P := \sum_{i\in\cT_P} \proj{e_I},
\end{equation}
(note that this is not uniquely specified by the distribution $P$
alone, but also requires specification of the basis $\ket{e_i}$), and
\begin{equation}
\Pi_{\rho,\delta} := \sum_{i\in\cT_{r,\delta}} \proj{e_I}.
\end{equation}
Statements on the cardinality of sets translate into statements on
the dimension of the corresponding subspaces (i.e. rank, or
equivalently, trace, of the projectors).
\par
Similarly, if we have states $W_i$ with diagonalizations
$W_i=\sum_j W(j|i)\proj{e_{j|i}}$, we can define, for any subset
$\cA\subset\cJ^n$ and $I\in\cI^n$
\begin{equation}
\Pi_{\cA}(I) := \sum_{J\in\cA} \proj{e_{J|I}}.
\end{equation}
This leads to the concept of \emph{conditional typical subspace projector},
for $\delta\geq 0$,
\begin{equation}
\Pi_{W,\delta}(I) := \sum_{J\in\cT_{W,\delta}} \proj{e_{J|I}},
\end{equation}
and again probability and cardinality statements about the typical
sequences translate into equivalent statements about certain traces.
\par
In particular we shall use the following estimate of the rank of the
conditional typical subspace projector:
\begin{equation}
\label{eq:conditional:typical:subspace:upper}
\Tr\Pi_{\rho,\d}(I) \leq
     (n+1)^{|\cI|d}\exp\bigl( n(S(\rho|P_I)+|\cI|d\eta(\d|\cI|/\sqrt{n})) \bigr).
\end{equation}
(Here we make use of the notation $S(\rho|P_I) := \sum_i S(W_i)$ in an
attempt to match the statements about typical sequences as closely as
possible.)  We'll also use the important probability estimate
\begin{equation}
\label{eq:weak:law:proj}
  \Tr\bigl(W_I\cT_{W,\d}(I)\bigr) \geq 1-\frac{|\cI|}{\d^2}.
\end{equation}

\subsection{Trade--off coding}\label{pfachieve}
We will use the coding technique that is summarized in the
following proposition. The statement is slightly more technical
and the estimates more explicit than we would need to prove our
main theorem~\ref{thm:central}. This is because we will re--use it
in section~\ref{sec:avs} and in section~\ref{sec:infinitesource}.
\begin{proposition}
\label{prop:coding} For a probability distribution $p$ on $\cI$ and a
classical noisy channel $p(\cdot|\cdot):\cI\rightarrow\cJ$ consider
the tripartite state $$\rho=\sum_i p_i\proj{i}^A \ox
\proj{\varphi_i}^B \ox \sum_j p(j|i)\proj{j}^C.$$ Then there exists a
visible code $(E,D)$ such that $$\forall I\in\cT_{p,\delta}\qquad
F\bigl( \proj{\varphi_I},(D\circ E)(I) \bigr)
                                                       \geq 1-\frac{4|\cI||\cJ|}{\delta^2}.$$
and having classical and quantum resources
\begin{eqnarray*}
  nS(A:C)  + nK|\cI||\cJ|\eta(\d/\sqrt{n}) +K'|\cI||\cJ|\log(n+1)
                                                              & & \text{classical bits}, \\
  nS(A:B|C)+ n\cdot 3d|\cI||\cJ|\eta(2\delta|\cI||\cJ|/\sqrt{n}) +  d|\cJ|\log(n+1)
                                                              & & \text{quantum bits},
\end{eqnarray*}
where $K$ and $K'$ are absolute constants.
\end{proposition}
\begin{proof}
  We design an $n$--block code as follows (typicality conditions throughout are
  with respect to a previously fixed $\delta$):
  \begin{itemize}
    \item Encoding:\\
      1.\, Given $I$ generate $J$ according to $p(J|I)$.\\
        2.\, Compress (i.e., project) the quantum state $\proj{\varphi_I}$
          to the conditional typical subspace
          $\Pi_{\widetilde{\rho}^{IJ},\delta}(J)$,
          where $\widetilde{\rho}^{IJ}_j=\sum_i
          W_{I|J}(i|j)\proj{\varphi_i}$.\\[1mm]
      If $I$ is typical and $J$ is conditionally typical, send $J$ and the
      joint type of $I$ and $J$ as classical data, and send the
      projected state on
       $\Pi_{\widetilde{\rho}^{IJ},\delta}(J)$ as quantum data.
    \item Decoding:\\ Given $J$, one can isometrically embed the quantum state transmitted
      back into the ambient Hilbert space.
  \end{itemize}
  The fidelity of this scheme is analyzed as follows. (We assume that if,
  at any point of the above protocol, an ``if'' is not satisfied, then some
  fixed failure action is taken.  Such would be the case when the POVM
  involving the above subspace projection yields an orthogonal result,
  for example.)
  With probability at least $1-|\cI|/\d^2$, $J$ is conditionally typical,
  and in this case the projection is successful with probability
  at least $1-|\cJ|/\d^2$ (by virtue of Eq.~(\ref{eq:weak:law:proj})),
  leaving a state which (cf. ~\cite{schumacher}) has fidelity
  $\geq 1-2|\cJ|/\d^2$ to $\proj{\varphi_I}$.
  \par
  Looking at the classical cost of this procedure, we see that it is
  dominated by sending $J$, which requires too many, namely
  $nS(C)$, classical bits. Here the Reverse Shannon
  Theorem~\cite{bennett:et:al} is invoked.  (For a precise statement,
  see theorem~\ref{thm:rst} below.)
  Using this theorem we can simulate the channel $p$ on the typical sequences $I$
  sending $nS(A:C)+o(n)$ classical bits, but at the same time needing
  an amount of shared randomness. The simulation, in fact, has the
  property that it endows sender and receiver with a common $J$, the
  distribution of which is $\frac{|\cI||\cJ|}{\d^2}$--close to $p(J|I)$.
  Taking all these points into account, we see that the fidelity of this
  protocol is at least $1-\frac{3|\cI||\cJ|}{\d^2}$ for every individual
  $\proj{\varphi_I}$ for which $I$ is typical.
  \par
  The analysis of the quantum resources needed is equally straightforward.
  By Eq.~(\ref{eq:conditional:typical:subspace:upper})
  the number of qubits needed to transmit the projected state is
  \begin{equation}
    \label{eq:qubit:resources}
    nS(\widetilde{\rho}^{IJ}|P_J) + dn|\cJ|\eta(\d/\sqrt{n}) + d|\cJ|\log(n+1).
  \end{equation}
  Note that the leading term is a conditional von Neumann entropy
  of the bipartite state
  \begin{equation}
  \rho  =\sum_j \widetilde{\rho}^{IJ}_j \ox P_J(j)\proj{j},
  \end{equation}
  which has trace norm distance at most $2\d|\cI||\cJ|/\sqrt{n}$ from
  \begin{equation}
  \omega=\sum_{ij} p(i)\proj{\varphi_i} \ox p(j|i)\proj{j}.
  \end{equation}
  (This follows from the typicality of $I$ and conditional typicality
  of $J$.)
  Next using the Fannes inequality~(\ref{eq:fannes}),
  we can upper bound Eq.~(\ref{eq:qubit:resources}) by
  \begin{equation}
  nS(\widetilde{\rho}|q) + 2dn|\cJ|\eta(2\d|\cI||\cJ|/\sqrt{n})
                                  + dn|\cJ|\eta(\d/\sqrt{n}) + d|\cJ|\log(n+1),
  \end{equation}
  with $q_j=\sum_i P(i)p(j|i)$ and
  $\widetilde{\rho}_j=q_j^{-1}\sum_i P(i)p(j|i)\proj{\varphi_i}$.
  \par
  We are left with one remaining feature to address: the protocol uses shared
  randomness (and to a considerable extent,
  according to theorem~\ref{thm:rst}).
  We shall now show that we can reduce this requirement to
  $O(\log n)$ shared random bits using a technique very much like
  the derandomization argument in \cite{derandomization}. 
  The proof will then be complete because setting up these bits
  can be absorbed into the classical communication with
  asymptotically vanishing cost per letter.
  (Actually, in order to achieve 
  high average fidelity, no random bits are needed at all but our goal is to
  prove that high fidelity can be achieved for every state in the typical
  subspace, a more stringent requirement that is used later in our
  study of arbitrarily varying sources.)  
  \par
  Observe that a protocol using shared randomness can be viewed as a
  probabilistic mixture of ordinary, deterministic protocols.
  Index these by a variable $\nu$,
  accompanied by a probability $x_\nu$. For each $\nu$
  we have a corresponding fidelity $F_I(\nu)$ for each individual $I$.
  Our construction shows that for typical $I$,
  \begin{equation}
  \sum_\nu x_\nu F_I(\nu) \geq 1-\frac{3|\cI||\cJ|}{\d^2}=:\mu.
  \end{equation}
  Note that the left hand side is exactly the expectation of the
  random variable $F_I$. We now choose $\nu_1,\ldots,\nu_L$
  independently and identically distributed (i.i.d.), according to the
  probabilities $x_\nu$. For fixed $I$ the $F_I(\nu_l)$, $l=1,\ldots,L$
  are i.i.d. as well, and in the interval $[0,1]$. Thus we can apply
  the Chernoff--Hoeffding bound for their arithmetic mean
  (lemma~\ref{lemma:chernoff} below):
  \begin{equation}
  \Pr\left\{ \frac{1}{L}\sum_{l=1}^L F_I(\nu_l) < (1-\e)\mu \right\}
                                 \leq \exp\left( -L\frac{\e^2\mu}{2\ln 2} \right).
  \end{equation}
  By the union bound we can estimate the probability that the above event
  occurs for a single typical $I$ to be less than or equal to
\begin{equation}
\exp\left( -L\frac{\e^2\mu}{2\ln 2} \right) |\cI|^n.
\end{equation}
  Choosing $\e=|\cI||\cJ|/\d^2$, this bound is itself less than $1$ if
\begin{equation} \label{eqn:Lconstraint}
  L > \frac{2\d^4\ln 2}{|\cI|^2|\cJ|^2\mu} n\log|\cI|,
\end{equation}
  in which case we can conclude that there exist values $\nu_1,\ldots,\nu_L$
  such that, for all typical $I$, we have
  $$\frac{1}{L}\sum_{l=1}^L F_I(\nu_l) \geq 1-\frac{4|\cI||\cJ|}{\d^2}.$$
  Therefore, a shared uniform distribution over the numbers $1,\ldots,L$ is
  sufficient, where $L$ need only satisfy Eq.~(\ref{eqn:Lconstraint}).
  This can be accomplished with $O(\log n)$
  shared random bits, which is what we wanted.
\end{proof}
\par\medskip
Here are the auxiliary results we needed in the proof:
\begin{theorem}[Reverse Shannon Theorem. See~\cite{bennett:et:al} and~\cite{p:d:compress}]
  \label{thm:rst}
  For any channel $W:\cI\rightarrow\cJ$, distribution $P$ on
  $\cI$, and $0<\lambda<1$ there exist maps
  \begin{align*}
    E_\nu &:\cI^n     \longrightarrow \{1,\ldots,M\} ,\\
    D_\nu &:\{1,\ldots,M\} \longrightarrow \cJ^n,
  \end{align*}
  $\nu=1,\ldots,N$, such that
  $$\forall I\in\cT_{P,\delta}\qquad
             \frac{1}{2}\left\| W(I)-\frac{1}{N}\sum_{\nu=1}^N D_\nu(E_\nu(I)) \right\|_1
                                                               \leq \frac{|\cI||\cJ|}{\d^2}.$$
  Moreover, with an absolute constant $K$,
  \begin{align*}
    \log M &\leq nH(P:W)+ nK|\cI||\cJ|\eta(\d/\sqrt{n}) +K|\cI||\cJ|\log(n+1), \\
    \log N &\leq nH(W|P)+ nK|\cI||\cJ|\eta(\d/\sqrt{n}) +K|\cI||\cJ|\log(n+1).
  \end{align*}
  \qed
\end{theorem}
\begin{lemma}[Chernoff--Hoeffding bound~\cite{chernoff,hoeffding}]
\label{lemma:chernoff}
Let $X_1,\ldots,X_L$ be independent, identically distriuted random variables,
taking real values in the interval $[0,1]$, and with expectation $\EE X_l\geq \mu$.
Then, for $\epsilon>0$,
$$\Pr\left\{ \frac{1}{L}\sum_{l=1}^L X_l < (1-\e)\mu \right\}
                             \leq \exp\left( -L\frac{\e^2\mu}{2\ln 2} \right).$$
\qed
\end{lemma}
\par
With this we are ready to state our main result:
\begin{theorem}
  \label{thm:central}
  $Q^*(R)=M(\cE,R)$.
\end{theorem}
\begin{proof}
  The inequality ``$\geq$'' is theorem~\ref{thm:lowerbound}.
  For the opposite inequality choose a $p(\cdot | \cdot)$ such that
  $S(A:C)\leq R$ and $S(A:B|C)\leq M(\cE,R)+\e$.
  Then, according to proposition~\ref{prop:coding}, there exist
  $n$--block codes $(E,D)$ with classical and quantum rates bounded
  by $R+o(1)$ and $M(\cE,R)+\e+o(1)$, respectively, which
  have fidelity $1-\e$ for all \emph{typical} $I$. But since these
  carry almost all the probability weight (say, larger than $1-\e$)
  of all sequences, the fidelity of the scheme is at least $1-2\e$,
  regardless of what is done on non--typical sequences.
  As $\e$ was arbitrary, we get $Q^*(R)=M(\cE,R)$.
\end{proof}
\medskip
\begin{remark}
The proof of proposition~\ref{prop:coding}, as the eventual
``derandomization'' shows, does not use the full power of the reverse
Shannon theorem, but only a consequence that is actually also used in
rate--distortion coding: that one can map the typical sequences $I$
onto $\exp(nH(P:W)+o(n))$ many $J$'s such that all the pairs
$(I,f(I))$ are jointly typical.
\end{remark}

\section{Exploring the trade--off curve} \label{sec:explore}
In this section we use our formula for the trade--off curve
to evaluate $Q^*(R)$ numerically for a selection of particular
ensembles chosen to illustrate further important properties of
the trade--off function.

\begin{figure}
\begin{center}
\epsfxsize=3.6in\epsfbox{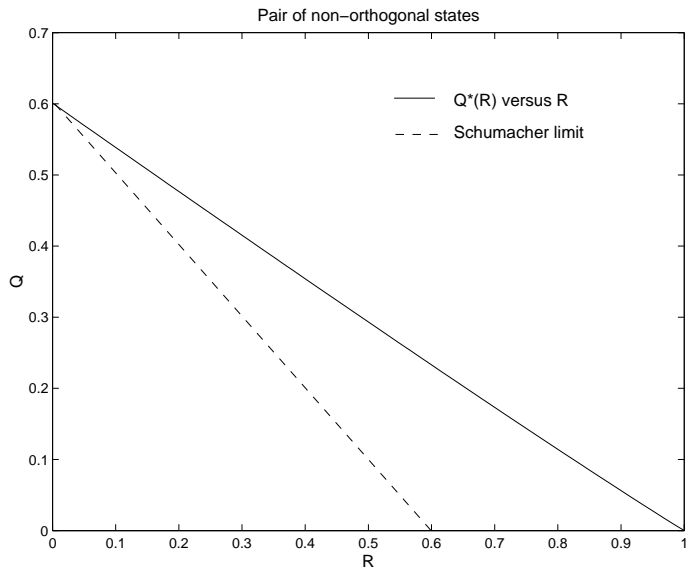}
\end{center}
\caption{The trade--off curve for a pair of equiprobable,
non-orthogonal states.  The dashed line represents the lower bound
$Q^*(R)+R \geq S(\cE)$ imposed by the Schumacher limit.}
\label{fig:pair:graph}
\end{figure}

To begin, let us consider the very simplest possibility, a pair of
non--orthogonal states.  Figure \ref{fig:pair:graph} plots the
trade--off curve for the pair $\{\ket{0},
\smfrac{1}{\sqrt{2}}(\ket{0}+\ket{1})\}$, each occurring with
probability $1/2$.  At first glance, $Q^*(R)$ appears to coincide with
the linear upper bound given by interpolating between $(0,S(\cE))$
and $(H_2(1/2),0)$.  A more detailed examination, however, reveals that
the curve is actually very slightly nonlinear.  Therefore, somewhat
surprisingly, the simple
quantum--classical coding scheme given by time-sharing between
fully quantum and fully classical coding is nearly optimal but
not completely so.  As we'll see below, this need not always be true.

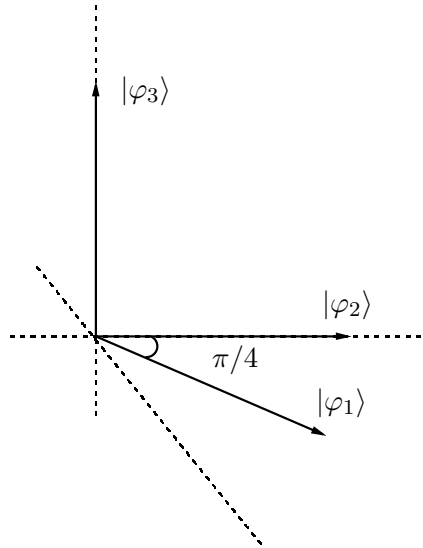
\begin{figure}
\begin{center}
\setlength{\unitlength}{0.00043333in}
{\renewcommand{\dashlinestretch}{30}
\begin{picture}(5049,6639)(0,-10)
\thicklines \put(1681.500,2440.500){\arc{267.152}{5.1409}{7.8877}}
\path(1062,2562)(4137,2562)
\blacken\path(4017.000,2532.000)(4137.000,2562.000)(4017.000,2592.000)(4017.000,2532.000)
\thinlines \dashline{60.000}(1062,1587)(1062,6612)
\dashline{60.000}(343,3412)(3087,12) \thicklines
\path(1062,2562)(3837,1362)
\blacken\path(3714.950,1382.094)(3837.000,1362.000)(3738.764,1437.165)(3714.950,1382.094)
\path(1062,2562)(1062,5637)
\blacken\path(1092.000,5517.000)(1062.000,5637.000)(1032.000,5517.000)(1092.000,5517.000)
\thinlines \dashline{60.000}(12,2562)(5037,2562)
\put(2487,2187){$\pi/4$} \put(1362,5487){$\ket{\ph_3}$}
\put(3837,2862){$\ket{\ph_2}$} \put(3762,1662){$\ket{\ph_1}$}
\end{picture}
}
\end{center}
\caption{The three--state ensemble $\cE_3$ consists of the states
$\ket{\ph_1}$, $\ket{\ph_2}$, $\ket{\ph_3}$ occurring with equal
probabilities.} \label{fig:3state:pic}
\end{figure}

\begin{figure}
\begin{center}
\epsfxsize=3.6in\epsfbox{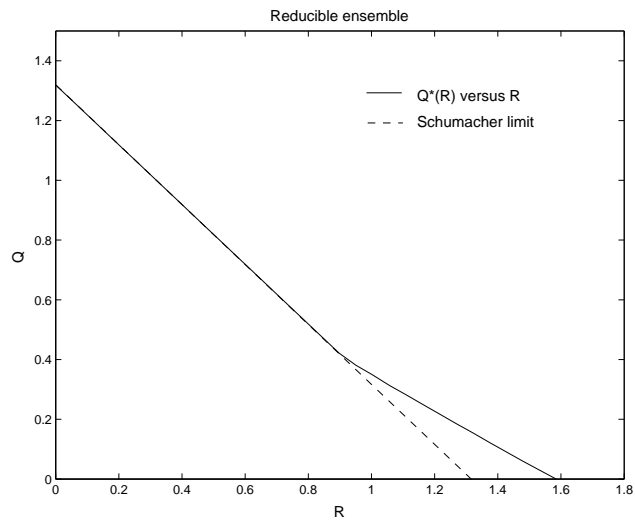}
\end{center}
\caption{ The trade--off curve for three--state ensemble $\cE_3$. The
dashed line again represents the Schumacher lower bound, which in
this case is achievable for $R \leq H(1/3)$.} \label{fig:3state:graph}
\end{figure}

In general, more complicated ensembles with internal structure will
have trade--off curves reflecting that structure. Consider, for
example, the three--state ensemble $\cE_3$ illustrated in figure
\ref{fig:3state:pic}, consisting of the states $\ket{\ph_1} = \ket{0}$,
$\ket{\ph_2}= \smfrac{1}{\sqrt{2}}(\ket{0}+\ket{1})$ and
$\ket{\ph_3}=\ket{2}$ with equal probabilities.
Since the set of states decomposes into two subsets $\cX_1 = \{
\ket{\ph_1},\ket{\ph_2} \}$ and $\cX_2 = \{ \ket{\ph_3} \}$ with
mutually orthogonal supports, it is possible to encode whether a given
$\ket{\ph_i} \in \cX_1$ or $\ket{\ph_i} \in \cX_2$ efficiently using
$H_2(1/3)$ classical bits.  Indeed, figure \ref{fig:3state:graph}
plots $Q^*(R)$ for this ensemble and we see that the Schumacher limit
is achieved for values of $R \leq H_2(1/3)$. For values of $R >
H_2(1/3)$, or once the classical information in the ensemble has
been exhausted, the trade--off curve departs from the Schumacher
lower bound to meet the point $(H(1/3,1/3,1/3),0)$.

\begin{figure}
\begin{center}
\epsfxsize=3.6in\epsfbox{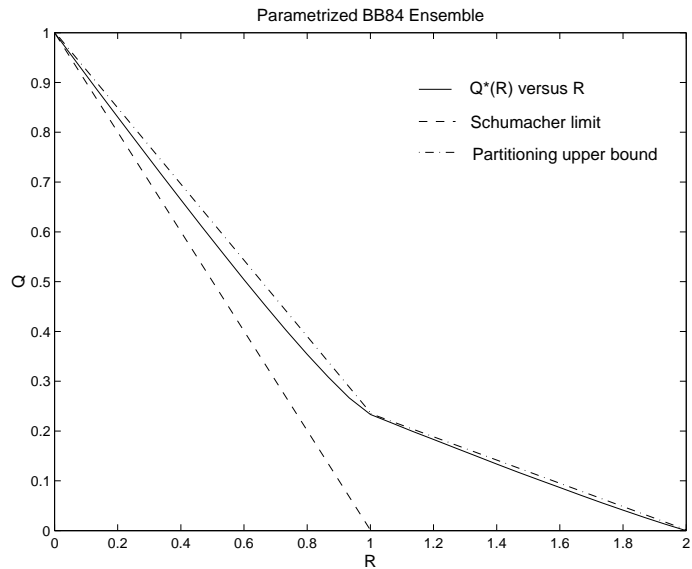}
\end{center}
\caption{Trade--off curve for the BB84 ensemble $\cE_{BB}(\pi/8)$.
The dashed line represents the Schumacher lower bound and the
dashed-dot line the upper bound from partitioning into the sets
$\cX_1$ and $\cX_2$.} \label{fig:BB84:graph}
\end{figure}

Our third example, the parametrized BB84 ensemble $\cE_{BB}(\t)$
introduced in section \ref{sec:compression}, is an ensemble that,
like $\cE_3$ above, decomposes naturally into subensembles.  On the
other hand, unlike for $\cE_3$, the subensembles are generally not
orthogonal.  The trade--off curve for $\t=\pi/8$ is plotted
in figure \ref{fig:BB84:graph}.  As usual, the dashed lower bound is
the Schumacher limit.  The dashed--dot line is the piecewise linear
upper bound constructed in section \ref{sec:compression}.
Squeezed into the intermediate region, we see that
$Q^*(R)$ is typically strictly less than the upper bound and,
especially in the region $0 < R < 1$, quite strongly curved.  The
point $(1,H_2(\smfrac{1}{2}(1 + \cos \smfrac{\pi}{8}))$ provides another
surprise: $Q^*(R)$ and the upper bound coincide there.  Therefore,
the partitioning scheme is optimal if exactly one bit of classical
storage is to be consumed per copy but not otherwise.
\par
\begin{figure}[t]
\begin{center}
\begin{psfrags}
\psfragscanon
\psfrag{con1}[l][l]{\tiny{$M(\cE_1,R)$}}
\psfrag{con2}[l][l]{\tiny{$M(\smfrac{1}{2}\cE_1+\smfrac{1}{2}\cE_2,R)$}}
\psfrag{con3}[l][l]{\tiny{$\smfrac{1}{2}M(\cE_1,R)+\smfrac{1}{2}M(\cE_2,R)$}}
\psfrag{con4}[l][l]{\tiny{$M(\cE_2,R)$}}
\epsfxsize=3.6in\epsfbox{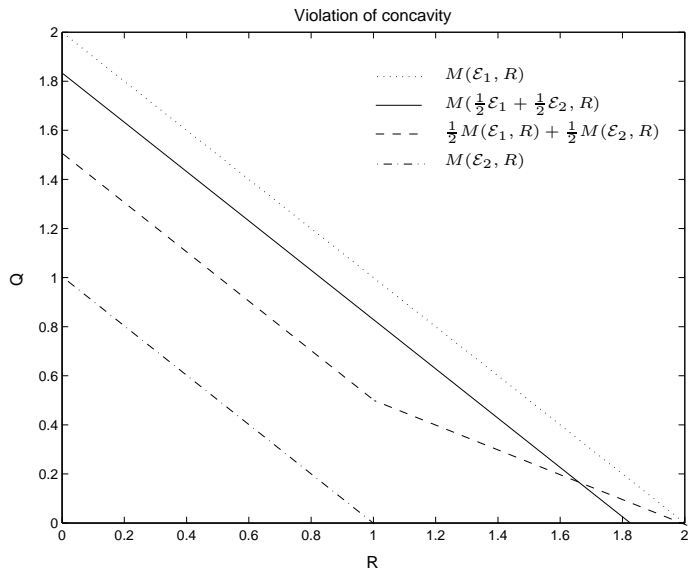}
\end{psfrags}
\end{center}
\caption{Violation of concavity in the ensemble.  If $Q^*$ were concave
in the ensemble, the solid line representing $M(\smfrac{1}{2}\cE_1+\smfrac{1}{2}\cE_2,R)$ would always exceed the dashed line of
$\smfrac{1}{2}M(\cE_1,R)+\smfrac{1}{2}M(\cE_2,R)$.  For large values of
$R$ we see that is not the case in this example.}
\label{fig:not:concave}
\end{figure}

\medskip
We now turn to another interesting property of the trade--off curve.
Contrary to what one might expect, the function $M(\cE,R)$ is
\emph{not concave in the ensemble}, violating the intuition that it
should be harder to send the mixture of two ensembles than it is to
probabilistically send either one. (Note that $M(\cE,0)$, however, is
just the von Neumann entropy $S(\cE)$ and is, therefore, concave
in $\cE$.) In fact, counterexamples to concavity can be constructed
without even making use of nonorthogonal states.  Let
$\cE_1 = \{ \ket{i},1/4 \}_{i=0}^3$ be an ensemble consisting
of four equiprobable orthonormal states and let
$\cE_2 = \{ \ket{i}, 1/2 \}_{i=0}^1$.  We can also consider the
mixture of ensembles
\begin{equation}
\cE := \smfrac{1}{2} \cE_1 + \smfrac{1}{2} \cE_2
= \{ (\ket{0},3/8),(\ket{1},3/8),(\ket{2},1/8),(\ket{3},1/8) \}.
\end{equation}
Since each of these ensembles is effectively classical, the
Schumacher lower bound is attainable and their trade-off
curves are just straight lines with slope $-1$.  From there, we can also
evaluate $\smfrac{1}{2}(M(\cE_1,R)+M(\cE_1,R))$ and compare it
to $M(\cE,R)$.  This is done in figure \ref{fig:not:concave},
revealing a violation of concavity when $R$ comes close to $2$.

%Since the reason for the violation of concavity is
%general, we will construct a family of examples and then specialize
%to a particular pair of ensembles:
%\par
%For ensembles $\cE_1,\cE_2$, assumed to be orthogonal to each other,
%it is easy to see that (with $\l_1+\l_2=1$)
%\begin{equation}
%\label{eq:mixture} M(\l_1\cE_1+\l_2\cE_2,R) \leq H(\l_1,\l_2) +
%\sum_k \l_k M(\cE_k,R_k),
%\end{equation}
%where $\l_1 R_1+\l_2 R_2 = R$. Then a counterexample to
%$$M(\l_1\cE_1+\l_2\cE_2,R) \geq \l_1 M(\cE_1,R)+ \l_2 M(\cE_2,R)$$
%can be constructed by favouring very different values of $R_1$ and
%$R_2$: for example, for $R_2=0$, Eq.~(\ref{eq:mixture}) can be
%estimated
%\begin{equation}
%M(\l_1\cE_1+\l_2\cE_2,R) \leq 1+\l_2 S(\cE_2)+\l_1 M(\cE_1,R/\l_1).
%\end{equation}
%Choosing $\l_1=\l_2=1/2$ and $\cE_2$ some fixed ensemble, we only
%need to find $\cE_1$ such that
%\begin{equation}
%M(\cE_1,2R) < M(\cE_1,R) - \bigl( 2+S(\cE_2) \bigr).
%\end{equation}
%One such pair of ensembles $\cE_1$ and $\cE_2$ is given by (...). The
%trade--off curve $M\bigl(\smfrac{1}{2}(\cE_1+\cE_2)\bigr)$ is plotted against
%$\smfrac{1}{2}\bigl(M(\cE_1,R)+M(\cE_2,R)\bigr)$ is plotted in figure
%\ref{fig:not:concave}. A clear violation of concavity is visible in
%the region (...).
\par
In the same spirit, note that an analogous construction shows that, while
\begin{equation}
M(\cE_1\ox\cE_2,2R) \leq M(\cE_1,R)+ M(\cE_2,R)
\end{equation}
always holds, equality (i.e., the natural ``additivity'' property of
$M$ under tensor products) may be violated if the ensembles are
sufficiently different from each other. More generally we have:
\begin{proposition} \label{prop:tensorM}
\[ M(\cE_1\ox\cE_2,R) = \min \{M(\cE_1,R_1)+M(\cE_2,R_2) : R_1+R_2=R\}.
\]
\end{proposition}

Also, while $M(\cE,R)$ may not be concave in the ensemble $\cE$, it does
obey a weaker condition analogous to Schur concavity.
\begin{proposition} \label{prop:SchurConcave} Let $\cE = \{ \ket{\ph_i}, p_i
\}$ be an ensemble. Let $\{ a_k \}$ be a set of probabilities with
corresponding unitary operators $U_k$ and $\cF$ be the ensemble
 $\cF = \{ U_k \ket{\ph_i}, p_i a_k \}$. Then
$M(\cE,R) \leq M(\cF,R)$.
\end{proposition}
The proofs of these propositions can be found in the appendices
\ref{appendix:tensorM} and \ref{appendix:SchurConcave}, respectively.

\begin{figure}
\begin{center}
\epsfxsize=3.6in\epsfbox{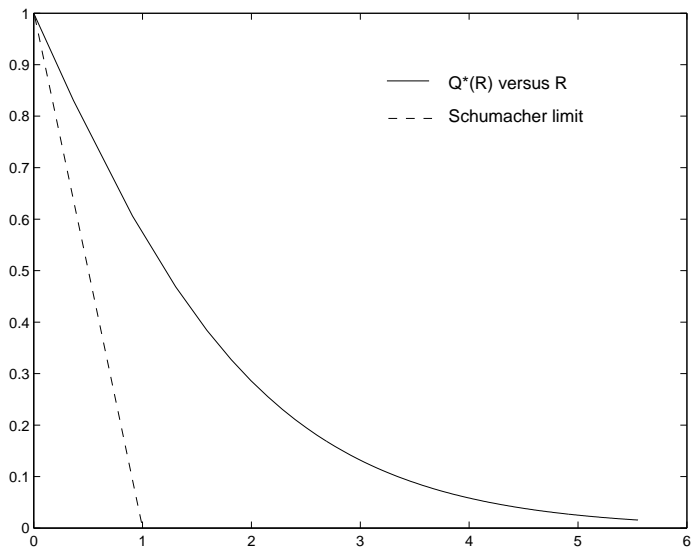}
\end{center}
\caption{Trade--off curve for the uniform qubit ensemble.  Note
that the curve never reaches the $Q=0$ axis, encoding the fact
that no finite amount of classical information is sufficient to
perfectly transmit an arbitrary qubit state.}
\label{fig:uniform:graph}
\end{figure}

As our last example, we include the trade--off curve for the uniform
(unitarily--invariant) ensemble on a single qubit as
figure~\ref{fig:uniform:graph}.  Devetak and Berger \cite{devetak:berger}
actually calculated an explicit parameterization of the optimal trade--off
curve for a restricted class of encodings.  Our lower bound of
theorem~\ref{thm:lowerbound}, or, rather, its infinite source ensemble
variant, theorem~\ref{thm:lowerbound:infinite}, proves that their
construction is optimal within all possible quantum--classical coding
strategies.  Thus, we can quote their result that, for $\l \in (0,\infty)$,
\begin{eqnarray}
R &=& \frac{\l}{\e^\l - 1} - 1 + \log\left(\frac{\l e^\l}{\e^\l -1}\right) \\
Q^*(R) &=& H_2\left(\frac{1}{\l}-\frac{1}{e^\l-1}\right)
\end{eqnarray}
gives a parameterization of $Q^*(R)$.
This curve will also play an important
role when we construct a probability--free version of our main result
in section \ref{sec:avs}.  We will find that, in an extremely strong
sense, it describes the cost of a qubit in classical bits.

\section{Arbitrarily varying sources} \label{sec:avs}
Our main result does not yet say, however, what a qubit is ``worth'' in bits
because it only supplies the trade--off curve $Q^*(R)$ for a given
set of quantum states once a set or prior probabilities have been
prescribed.  Without the probabilities, the curve is undefined and
the rate of exchange between bits and qubits can't be uniquely identified.
However, using the theory of \emph{arbitrarily varying sources (AVS)}
(see \cite{avs} for an exposition of this concept in classical
information theory), we can develop a probability--independent version
of our trade--off curve that will eliminate the ambiguity.
\par
Throughout this section, let $\cE$ denote not an ensemble, but just a set of
states, and let ${\bf P}\subset{\cal P}_\cE$ be a subset of
probability distributions on $\cE$. For each string $I\in\cI^n$ of
length $n$ we will consider product distributions
\begin{equation}
p^n(I) := p_1(i_1)\cdots p_n(i_n)
\end{equation}
where each $p_k \in {\bf P}$.
An \emph{AVS--code of fidelity} $1-\e$ is defined as a
visible code, as before (see definition~\ref{Cdefn:quantumEncDec}),
only that now the fidelity condition is required to hold for all
probability distributions in ${\bf P}$:
\begin{equation}
\label{eq:AVS:condition}
\forall p^n\in{\bf P}^n\qquad \sum_I p^n(I) F\bigl( \proj{\phi_I},(D\circ E)(I) \bigr)
                                                                       \geq 1-\epsilon.
\end{equation}
The classical and quantum rates are exactly as in
definition~\ref{defn:quantumResource} and likewise,
definition~\ref{Cdefn:quantumResourceCompression} can be used
unchanged to characterize attainable rate pairs $(R,Q)$. This leads
to the definition of the trade--off function $Q^*(R,{\bf P})$ as the
minimum $Q$ such that $(R,Q)$ is attainable.
\par
Intuitively, the encoder--decoder pair plays a game against a
clairvoyant adversary whose aim is to minimize their average fidelity
and who can control the source mechanism so as to
create any of the distributions $p^n\in{\bf P}^n$.
Their goal is to win by keeping the average
fidelity above $1-\epsilon$ against arbitrary strategies of the
adversary.
\par
A special case is that of ${\bf P}={\cal P}_{\cE}$, in which case
we have no restriction on the source, so that all possible state
strings are to maintain high fidelity.
\par
We shall use the notation $M(\cE,p,R)$ to designate our earlier
function $M$ for the ensemble consisting of the states $\cE$ and
the probabilities $p$, and define now
\begin{equation}\label{eqn:tradeoffAVS}
M(\cE,{\bf P},R) := \sup_{p\in{\bf Q}} M(\cE,p,R)
\end{equation}
where ${\bf Q}:={\rm conv}({\bf P})$ is the convex hull of ${\bf P}$.
\begin{theorem}
\label{thm:AVS:tradeoff}
$Q^*(R,{\bf P})=M(\cE,{\bf P},R)$.
\end{theorem}
\begin{proof}
The inequality ``$\geq$'' follows almost directly from
theorem~\ref{thm:lowerbound}: only observe that the adversary can
simulate any source ensemble $p\in{\bf Q}$, and then
theorem~\ref{thm:lowerbound} applies. (More formally, choose a
probability distribution $s$ on ${\bf P}$ such that $p=\sum_k s_k
p_k$, and note that averaging Eq.~(\ref{eq:AVS:condition}) over
the measure $s^{\ox n}$ gives (\ref{eq:AVS:condition}) for $p^{\ox
n}$.)
\par
In the other direction, we only need to exhibit a covering of the
union of the ``probable sets'' of the distributions $p^n\in{\bf
P}^n$ by appropriate sets of typical sequences,and apply
proposition~\ref{prop:coding}. This is done as follows:
\par
For $p^n=p_1\ox\cdots\ox p_n\in {\bf P}^n$ observe that the set
\begin{equation}
\cT_{p^n}:=\left\{ I : \forall i\ \left|N(i|I)-\sum_{k=1}^n p_k(i)\right|
                                                     \leq \delta\sqrt{n}\right\}
\end{equation}
carries (by Chebyshev's inequality) almost all the weight of the
distribution:
\begin{equation}
p^n(\cT_{p^n})\geq 1-\d^{-2}.
\end{equation}
Since $\cT_{p^n}$ is in fact the same as the set of typical
sequences $\cT_{\overline{p},\d}$, for
$\overline{p}=\frac{1}{n}\sum_k p_k\in{\bf Q}$, the union
$\bigcup_{p^n} \cT_{p^n}$ is actually a union of certain type
classes, and hence we may choose
$\overline{p}_1,\ldots,\overline{p}_T$, $T\leq (n+1)^{|\cI|}$,
such that
\begin{equation}
\cT:=\bigcup_{p^n\in{\bf P}^n} \cT_{p^n} = \bigcup_{t=1}^T \cT_{\overline{p}_t,\d}.
\end{equation}
\par
The coding is very simple: when $I\in\cT$ the encoder chooses $t$
such that $I\in\cT_{\overline{p}_t,\d}$. He then communicates $t$
to the decoder, and uses the protocol of
proposition~\ref{prop:coding}. (In fact, communication of $t$ is
not even necessary, as in the latter protocol the type of $I$ is
communicated anyway.) When $I\not\in\cT$ some fixed default choice
is sent.
\par
By construction and by proposition~\ref{prop:coding}, for
sufficiently large $\d$ this scheme uses $R+\e$ classical bits and
$M(\cE,{\bf P},R)+\e$ qubits per source symbol. For each
$p^n\in{\bf P}^n$ we obtain high fidelity for all states outside a
set of arbitralily small probability.
\end{proof}
\par\medskip
In particular, for the above--mentioned case of no restrictions at
all on the probabilities, we get the trade--off function
  \begin{equation}\label{eqn:tradeoffUnrestricted}
   {\bf Q^*}(R, \cP_{\cE})=\sup_{p\in\cP_{\cE}} M(\cE,p,R).
  \end{equation}
  which depends only on the states of $\cE$.
For a finite ensemble it is quite easy to show that $M(\cE,p,R)$
is continuous in the distribution $p$. This implies that the
suprema in Eqs. (\ref{eqn:tradeoffAVS}) and
(\ref{eqn:tradeoffUnrestricted}) are, in fact, \emph{maxima} (in
the former case over the closure of ${\bf Q}$).

\section{Information and disturbance} \label{sec:infoDisturb}
The function $M(\cE,R)$, in addition to providing the
quantum--classical trade--off curve, has a number of other useful
interpretations.  Recall from proposition \ref{prop:useEquality} that
\begin{eqnarray} \label{eqn:stdDefn}
M(\cE,R) &=& \inf_{p(\cdot|\cdot)} \{ S(A:B|C):S(A:C) = R \},
\end{eqnarray}
with an equality for $S(A:C)$ rather than the inequality we usually
use. By the chain inequality,
\begin{equation}
S(A:C)+S(A:B|C) = S(A:BC)
\end{equation}
and $S(A:BC)$ is just the Holevo $\chi$ quantity of the ensemble
\begin{equation}
\cF^{BC} := \{ \ph_i^B \ox \sum_j p(j|i) \proj{j}^C , p_i \}.
\end{equation}
Therefore, if we define the function $X(\cE,R):=R+M(\cE,R)$, then
we can re--write Eq. (\ref{eqn:stdDefn}) as
\begin{eqnarray}
X(\cE,R) = \inf_{p(\cdot|\cdot)} \{ \chi(\cF^{BC}) : S(A:C) = R \}.
\end{eqnarray}
The quantity on the right is now perhaps more familiar than the
conditional mutual information $S(A:B|C)$: it is a standard
measure of the distinguishability present in the ensemble
$\cF^{BC}$, minimized over all possible ways of including a fixed
amount of classical information about the index $i$ in register
$C$.  Now suppose that Alice is initially given a state
$\ket{\ph_i}$ from $\cE$ (without the name $i$ this time) and, via
a CPTP map, manages to extract an amount $R$ of classical
information about $i$ without damaging any of the states
$\ket{\ph_i}$.  Then her final Holevo $\chi$ would necessarily be
at least as large as $X(\cE,R)$, by definition. Typically,
however, $X(\cE,R) > S(\cE)$ (by the Schumacher lower bound to
$Q^*(R)=M(\cE,R)$), so such an operation will be forbidden by the
monotonicity of $\chi$. Therefore, it is impossible for Alice to
extract information without disturbing the states.

The simple argument above combined with the additivity of
$M_\e(\cE,R)$ from section \ref{subsec:mutualInfo} can be used to
prove interesting statements about the trade--off between
information gain and state disturbance in an asymptotic and
approximate setting. In contrast to the compression problem,
however, we can make stronger statements if we use the mean
letterwise fidelity measure $\lbar{F}$ section
\ref{subsec:alternativeDefs} instead of the global fidelity
measure $F$. Therefore, we will express our results in terms of
the corresponding function $\lbar{M}_\e(\cE^{\ox n},nR)$ instead
of $M_\e(\cE^{\ox n},nR)$.  Recall that these functions are
defined identically except that the first uses the mean fidelity
function $\lbar{F}$ and the second uses the global fidelity $F$.
Likewise, define $\lbar{X}_\e(\cE,R) = R + \lbar{M}_\e(\cE,R)$.
Since $F$ and $\lbar{F}$ are identical for a single copy, we have
$\lbar{M}_\e(\cE,R) = M_\e(\cE,R)$ and similarly for $X$ and
$\lbar{X}$. By the discussion in section
\ref{subsec:alternativeDefs}, we know that $\lbar{M}_\e(\cE^{\ox
n},nR) = n\lbar{M}_\e(\cE,R)$, which in turn implies
\begin{eqnarray} \label{eqn:barXadd}
\lbar{X}_\e(\cE^{\ox n},nR) = n X_\e(\cE,R).
\end{eqnarray}
Now, generalizing the above single copy argument, suppose that Alice
is given a state $\ket{\ph_I}$ drawn from $\cE^{\ox n}$, which by a
CPTP map $\G$, she manages to convert into the state
\begin{eqnarray}
\rho_I = \sum_j \tilde{\ph}_{I,j}^B \ox p(j|I) \proj{j}^C,
\end{eqnarray}
with a quantum and classical part such that the mutual information
$H(I:j) \geq nR$ and the mean letterwise fidelity between Alice's initial
states and her final states of system $B$ satisfies
\begin{eqnarray}
\lbar{F}\left(\cE^{\ox n},\Tr_C\circ \G(\cE^{\ox n})\right) := \sum_I
p_I \frac{1}{n} \sum_{k=1}^n F\left(\ph_{i_k},\Tr_{\neq k} \circ
\Tr_C(\r_I)\right) \geq 1-\e.
\end{eqnarray}
Writing $\cF^{BC} = \{ \G(\ph_I), p_I \}$, the monotonicity of $\c$
guarantees that $nS(\cE) \geq \c^{BC}$ and it is easy to see that
$\c^{BC} \geq \lbar{X}_\e(\cE^{\ox n},nR)$.  By applying
Eq.~(\ref{eqn:barXadd}), we then find
\begin{eqnarray}
S(\cE) \geq X_\e(\cE,R),
\end{eqnarray}
in which, conspicuously, all dependence on $n$ has vanished.  In
other words, in order to maximize her information at a given mean
letterwise fidelity, Alice should just repeat the optimal single
letter strategy for each position; she needn't ever apply any
collective operations.  Summarizing these observations, we have:
\begin{theorem}
Suppose we have a set of states $\ket{\ph_I}$ drawn from the
ensemble $\cE^{\ox n}$ represented on system $B$ and let $\G$ be a
CPTP map from $B$ to the joint system $BC$, where $C$ is classical,
satisfying the following conditions:
\begin{enumerate}
\item $H(I:j) \geq nR$, where $j$ is the classical output on system
$C$.
\item The mean letterwise fidelity
$\lbar{F}\left(\cE^{\ox n},\Tr_C\circ \G(\cE^{\ox n})\right) \geq 1 -
\e$.
\end{enumerate}
Then, for each $\e >0$, the inequality $S(\cE) \geq X_\e(\cE,R)$ holds.
Moreover, the Holevo quantity of the ensemble $\cF^{BC} = \{
\G(\ph_I), p_I \}$ satisfies the inequality $\c(\cF^{BC}) \geq n
X_\e(\cE,R)$. \qed
\end{theorem}
One application of the theorem is that it provides an alternative
method for analyzing the quantum resources required for blind
compression, which was the subject of \cite{BarnumHJW01}.  The
idea is simply to think of the map $\G$ as the composition $D_n
\circ E_n$ of the encoding and decoding maps for blocks of size
$n$.  (Because classical information can be copied, we can assume
without loss of generality that the decoder keeps his classical
information around after the decoding stage has been completed.)
Now suppose that the scheme has classical mutual information
$H(I:j) \geq nR$. If it also has mean letterwise fidelity $1-\e_n$
then, as for the visible case,
\begin{eqnarray}
\qsupp \geq \smfrac{1}{n} \lbar{M}_{\e_n}(\cE^{\ox n},nR) =
M_{\e_n}(\cE,R).
\end{eqnarray}
By the previous theorem, however, we must also have the inequality
$S(\cE) \geq X_{\e_n}(\cE,R)$.  Moreover, if perfect compression is
possible asymptotically (using either the block or letterwise
fidelity conditions), we get the stronger inequality
\begin{eqnarray} \label{eqn:chiConstraint}
S(\cE) \geq \lim_{\e\dn 0} X_{\e}(\cE,R) = X_0(\cE,R).
\end{eqnarray}
(The continuity at $\e=0$ follows from the continuity of $M_0$,
demonstrated earlier.)  Because the ensemble $\cE$ can always be
recovered by tracing over the $C$ register, the monotonicity of
$\chi$ guarantees that the right hand side is always at least as
large as the left, implying $S(\cE) = X_0(\cE,R)$.  We are,
therefore, interested in the equality conditions for monotonicity.

Recalling some terminology from \cite{BarnumHJW01}, we say an
ensemble $\cE$ is \emph{reducible} if its states can be partitioned
into two non--empty sets with orthogonal supports.  An ensemble is
said to be irreducible if it is not reducible.  Every ensemble,
therefore, can be decomposed into orthogonal, irreducible
subensembles as
\begin{equation}
\cE = \bigcup_{l=1}^L a_l \cE_l,
\end{equation}
where $a_l$ is the total probability weight of states in subensemble
$\cE_l$.
\begin{proposition} \label{prop:equality:conditions}
Let $\cE = \cup_{l=1}^L a_l \cE_l$ be a decomposition of the
pure--state ensemble $\cE$ into irreducible sub--ensembles $\cE_l
= \{ \ket{\ph_{il}},p_{i|l} \}$ and let $\cF^{BC} = \{ \ph_{il}^B
\ox \o_{il}^C,a_l p_{i|l} \}$ be a bipartite extension of the
ensemble $\cE$.  Then $S(\cE) = \chi(\cF^{BC})$ if and only if
$\o_{il} = \o_{jl}$ for all $i$, $j$, and $l$.
\end{proposition}
A proof is given in appendix \ref{appendix:equality:conditions}.
The meaning of the proposition is essentially that the only
information that can be stored on register $C$ without increasing
$\chi$ is the classical information already present on register
$B$, so that $\o_{il}$ must be a function of $l$ alone. Therefore,
in order to satisfy Eq.~(\ref{eqn:chiConstraint}) it is necessary
that $R \leq H(a_1,\dots,a_L)$.  Conversely, provided the
inequality holds, it is possible to extract $R$ bits per signal
without disturbance at the encoding stage, at which point the
encoding scheme we used for visible compression can be used to
achieve the quantum rate $S(\cE) - R$.  Putting these observations
together, we obtain an alternative demonstration of the main
theorem of \cite{BarnumHJW01}:
\begin{theorem}
Let $\cE = \bigcup_{l=1}^L a_l \cE_l$ be a decomposition of the
ensemble $\cE$ into orthogonal, irreducible subensembles.  Then blind
compression of $\cE$ to $Q$ qubits per signal plus auxiliary
classical storage is possible if and only if
\begin{eqnarray}
Q \geq \sum_l a_l S(\cE_l) = S(\cE) - H(a_1,\dots,a_L).
\end{eqnarray} \qed
\end{theorem}
Thus, the techniques we have introduced to analyze the visible
compression problem provide a unified framework for analyzing blind
compression as well. In fact, we will see in the next section that
the trade--off curve for yet another related problem -- remote state
preparation -- can also be calculated using similar methods.

\section{Application to remote state preparation}
\label{sec:rsp}
Remote state preparation, introduced in \cite{rsp:prl} in work
motivated by a conjecture of Lo's~\cite{lo:rsp}), is very similar
to what we have considered here: it is a visible coding problem
for quantum states involving classical resources, in the form of
communication, and quantum resources, this time in the form of
entanglement. Furthermore, these two types of resources can be
traded against each other so it is natural to study the optimal
trade--off curve.
\par
Without giving formal definitions, let $E^*(R)$ be the minimum
rate of entanglement sufficient for a remote state preparation
protocol with classical rate $R$, such that the average fidelity
tends to $1$ with growing blocklength.
\par
Given that entanglement can be set up using quantum communication
at a cost of one qubit per ebit, and that, on the other hand,
quantum communication can be accomplished using
teleportation~\cite{teleportation} at a cost of two cbits and one
ebit per qubit, it is clear that coding methods for the one
problem immediately yield (possibly suboptimal) procedures for the
other. (In fact, by making use of quantum--classical trade--off
coding, this resulted in the ``cap--method'' of~\cite{rsp:prl},
which was further refined in~\cite{devetak:berger}.)
\par
In~\cite{rsp:BIG} a method of remote state preparation is developed
that works for visible coding of product states and is more efficient
than teleportation: we really need only to use \emph{one} cbit and
one ebit per qubit, asymptotically.
\begin{theorem}[See~\cite{rsp:BIG}]
  \label{thm:1and1per1}
  Given a finite set $\cX$ of states (density operators) on ${\cal K}$,
  there is a probabilistic exact (one--shot) remote state
  preparation protocol working for all states in $\cX$
  and with failure probability uniformly $\epsilon$,
  using a maximally entangled state
  $\ket{\Ph}$ on ${\cal K}\otimes{\cal K}$ and classical communication of
  a message out of
  $$M\leq 1+\frac{2\ln 2}{\epsilon^2}\log(2|\cX|\dim{\cal K})\dim{\cal K}.$$
  \qed
\end{theorem}
This leads immediately to
\begin{theorem}
  \label{thm:rsp}
  For the source $\cE=\{\ket{\ph_i},p_i\}$ of quantum states,
  if $R\geq 0$ and $Q=Q^*(R)$, then $E^*(R+Q)\leq Q$.
  \par
  As a consequence, we obtain:
  $$E^*(R)\leq N(\cE,R):=\min_{p(\cdot|\cdot)} \{ S(A:B|C) : S(A:BC)\leq R \},$$
  minimization over the same set of tripartite states as in the
  definition of $M$.
\end{theorem}
\begin{proof}
  We apply the above theorem~\ref{thm:1and1per1} to the space $\cK$ of
  \emph{encoded states} of an optimal trade--off coding using $R$ cbits and
  $Q$ qubits per source symbol, and to the set of all possible encoded states:
  note that $|\cX|\leq \bigl(|\cI||\cJ|\bigr)^n$.
  \par
  By that result, we need $Q$ ebits to do this, and an additional
  $Q+o(1)$ cbits to the $R$ cbits from the trade--off coding.
\end{proof}
In fact, in~\cite{rsp:BIG} it is shown, by methods very similar to
those in section~\ref{sec:lowerBound}, that the above estimate for
$E^*$ is in fact an equality, and that our AVS considerations also
carry over:
\begin{theorem}
For the state set $\cE$ and AVS ${\bf P}$, $$E^*(R,{\bf P}) =
\sup_{p\in{\bf Q}} N(\cE,p,R),$$ with ${\bf Q}={\rm conv}({\bf P})$.
\qed
\end{theorem}
\par\medskip
For ${\bf P}$ the set of all distributions on the pure states (as
indeed for any symmetric family of distributions) we can prove
symmetry results like those in the upcoming
section~\ref{sec:symmetry}, and arrive at the conclusion that the
\emph{absolute trade--off} between cbits and ebits in remote state
preparation is given by the curve $N(\cP(\cH),u)$, where $u$ is
the uniform (i.e. unitarily--invariant) measure on the set
$\cP(\cH)$ of all pure states on $\cH$. Devetak and
Berger~\cite{devetak:berger} arrived at a slightly different curve
as an upper bound to the true trade--off, starting from
$M(\cP(\cH),u)$ as we did, but employing teleporation instead of
the newer technique in theorem~\ref{thm:1and1per1}. For this
reason their conjecture that their bound is tight is not
correct.\\[5mm]

{\Large\bf \begin{center} PART II: \\ SOME FURTHER GENERALIZATIONS
\end{center}}

\section{Symmetry in the ensemble} \label{sec:symmetry}
Our formulas for the trade--off curve, both in the
known and arbitrarily varying source case, can be considerably
simplified, if there is symmetry in the set of states.
\par
Assume that there is a group $G$ acting on the labels $i$ of the
states by a projective unitary representation $U_g$
\begin{equation}
  \forall g\in G,i\in\cI \qquad \proj{\varphi_{gi}} = U_g\proj{\varphi_i}U_g^\dagger.
\end{equation}
(We will present the following arguments for a finite group, but
the same applies for compact groups: in fact, we only need the
existence of an invariant measure, see~\cite{halmos}.) The action
of $G$ on $\cI$ induces an action on the probability distributions
on $\cI$, in a natural way: if $p\in\cP(\cI)$ is a distribution,
then $p^g(i)=p(g^{-1}i)$ defines the translated distribution.
Assume now further that the arbitarily varying source ${\bf P}$ is
stable under this induced action:
\begin{equation}
\forall p\in{\bf P}\qquad p^g\in{\bf P}.
\end{equation}
(In the ``known source'' case, ${\bf P}=\{p\}$, this simply means
that $p(gi)=p(i)$ for all $i\in\cI$ and $g\in G$.)
\par
By the formula for the trade--off curve, Eq.~(\ref{eqn:tradeoffAVS}),
we may assume that ${\bf P}$ is convex. Letting
\begin{equation}
{\bf P}^G:=\{p\in{\bf P}:\forall g\in G\ p^g=p\},
\end{equation}
we can then prove
\begin{theorem}
\label{thm:symmetric:source}
For any $G$--invariant state set and AVS ${\bf P}$,
\begin{equation}
M(\cE,{\bf P},R) = M(\cE,{\bf P}^G,R).
\end{equation}
\end{theorem}
\begin{proof}
The l.h.s. is by definition greater than or equal than the r.h.s.
\par
For the opposite inequality we make use of the ``restricted
concavity'' given in proposition~\ref{prop:SchurConcave}, for the
rotations $U_g$ applied with equal probabilities to the ensemble
$(\cE,p)$ we get:
\begin{equation}
M\left( \bigcup_g U_g\cE U_g^\dagger,\frac{1}{|G|}\sum_g p^g,R
\right)
                     \geq \frac{1}{|G|} M(U_g\cE U_g^\dagger,p^g,R) = M(\cE,p,R).
\end{equation}
Note that $\frac{1}{|G|} \sum_g p^g \in {\bf P}^G$ and since the state
set is $G$ invariant we have $\bigcup_g U_g\cE U_g^\dagger = \cE$
which proves our claim.
\end{proof}
\par\medskip
If $G$ acts \emph{transitively}, this leads to a dramatic
simplification of the formula for the AVS trade--off curve
(theorem~\ref{thm:AVS:tradeoff}): in this case the only
$G$--invariant distribution is the uniform distribution, so from
theorem~\ref{thm:symmetric:source} we obtain:
\begin{corollary}
\label{cor:transitive:action} For an AVS $(\cE,{\bf P})$ with
transitive group action under which ${\bf P}$ is stable, (e.g. for
${\bf P}=\cP_{\cE}$), we have $$Q^*(R,{\bf P}) = M(\cE,u,R),$$ where
$u$ is the uniform distribution on $\cE$. \qed
\end{corollary}
\par
The particular example of $\cE$ being the set of all pure states
on $\cH$ and ${\bf P}$ being the set of all distribution on $\cE$,
is arguably the setting for \emph{the} trade--off between
classical and quantum bits: the trade--off coding becomes a
statement solely about states, with no mention of prior
probabilities. Of course we have not yet justified the application
of our results to infinite state sets. The corresponding more
involved treatment of the coding bounds will be given in
section~\ref{sec:infinitesource} below.
\par
Given this generalization to infinite state sets, we conclude that
the \emph{absolute trade--off} for pure states on $\cH$ is given
by $M(\cP(\cH),u)$, with the uniform (i.e., unitarily--invariant)
measure $u$ on the set $\cP (\cH)$ of all pure states. The
Devetak--Berger curve introduced earlier corresponds to the case
$\cH=\CC^2$.
\par\medskip

\begin{remark}
From the proof of theorem~\ref{thm:symmetric:source}, we see that we
may always restrict the classical encodings $p(\cdot|\cdot)$ to be
group covariant as well, in the sense that, for each $j\in\cJ$, the
distribution $q(\cdot|j)$ has the property that for each $g\in G$
there exists a $j'$ satisfying $q_{j'}=q_j$ and $q(gi|j)=q(i|j')$ 
for all $i\in\cI$:
\par
Define a new encoding $p'$ by letting
\begin{equation}
p'(j,g|gi):= \frac{1}{|G|} p(j|i).
\end{equation}
For a $G$--invariant distribution $p$ on the ensemble states this
does not change the values of $S(A:C)$ and $S(A:B|C)$. However,
the resulting probabilities $q^\prime_{j,g}=q_j$ and
$q'(gi|j,g)=p_i p(j|i)/{q^\prime_{j,g}}$ have a useful property:
there is a group action of $G$ on the indices $(j,g)$ under which
the distribution $q'$ is invariant, and the set of conditional
distributions $q'(\cdot|j,g)$ is stable. More precisely, $h$ acts
on $(j,g)$ by $h\cdot(j,g)=(j,hg)$. Obviously, $q'$ is invariant
under this, and
\begin{equation}
q'\bigl(gi|h\cdot(j,g)\bigr)=q'(gi|j,hg)=q'(h^{-1}hgi|j,gh),
\end{equation}
saying that
$q'\bigl(\cdot|h\cdot(j,g)\bigr)=\bigl(q'(\cdot|j,hg)\bigr)^h$.
\par
Hence, when discussing optimal codings given by $q_j$ and
$q(\cdot|j)$ such that $\sum_j q_j q(\cdot|j)=p$, we may always
assume that $G$ also acts on the set of $j$'s, and that
\begin{equation}
\forall j\forall g\qquad q_{gj}=q_j \text{ and }
q(\cdot|gj)=\bigl(q(\cdot|j)\bigr)^g.
\end{equation}
\end{remark}
\par\medskip
We close this section by giving a bound on the size of the classical
register for a finite ensemble with symmetry, which sometimes
improves our earlier result in proposition \ref{prop3.two}:
\begin{proposition}
\label{prop:small:classical:part} Let the group $G$ act on the
ensemble $\cE=\{\varphi_i,p_i\}_{i\in\cI}$ in the way described at
the beginning of this section, and assume that $p$ is $G$--invariant.
If the group action  partitions  $\cI$ into $t$ $G$--orbits then for
every $R$ there exists a classical encoding
$p(\cdot|\cdot):\cI\longrightarrow\cJ$ which is covariant in the
above sense, and satisfies $$|\cJ|\leq |G|(t+1),\qquad S(A:C)\leq
R,\qquad S(A:B|C)=M(\cE,R).$$
In fact, $\cJ$ partitions into $t+1$ $G$--orbits, in the sense 
described above.
\end{proposition}
The proof is given in Appendix \ref{pfprop:small}

\begin{example}
Let $\cE$ consist of any two states: $\cE = \{ \ket{\ph_i} \}_{i=1}^2$.
By choosing a reflection that swaps $\ket{\ph_1}$ and $\ket{\ph_2}$, 
we get a transitive $\ZZ_2$ action on the indices $i$.  Therefore, for 
the AVS $(\cE,\cP_\cE)$, we have $Q^*(R,{\bf P})=M(\cE,u,R)$, where $u$
is the uniform distribution $p_i = 1/2$.  This distribution is
clearly $G$-invariant so 
proposition \ref{prop:small:classical:part} ensures that there is
an optimal encoding for which $\cJ$ partitions into at most $t+1=2$ 
orbits, each of size either $1$ or $2$.
\end{example}

\begin{example}
For states in the BB84 ensemble $\cE_{BB}(\t)$, the group $\ZZ_2 \times \ZZ_2$ 
acts transitively via reflection along the $\t/2$ axis and rotation
by $\p/2$.  Therefore, once again, the unrestricted AVS can be reduced
to the uniform ensemble, for which the optimal encoding can be assumed
$G$-covariant, with $\cJ$ partitioning into at most two orbits of length
$1$, $2$ or $4$.
\end{example}

\section{Infinite source ensembles}
\label{sec:infinitesource} It should be noted that, even in the
technical parts of our proofs, and, indeed, in the very statements of
the \emph{coding theorems}, we assumed that the sets of states
under consideration were \emph{finite}.
\par
As there are interesting examples of ensembles with infinite state
sets, including perhaps most notably the whole manifold of pure
states in a Hilbert space, we show here how a certain
approximation technique (used in~\cite{winter:diss} to deal with
coding for nonstationary quantum channels) can be used to transfer
our main results quite directly. The procedure, unfortunately, is
not entirely painless; we have to go through the proof of
proposition~\ref{prop:coding} again with a modified and more
technical version of the typical subspace. That is why we have
chosen to treat the infinite source case separately, confining the
details to this section.

\subsection{Formulation of information quantities and the lower bound}
\label{subsec:measure} To be able to consider infinite ensembles and
encodings, we have to reformulate our notions from
sections~\ref{sec:compression} and~\ref{sec:lowerBound} in terms of
general measure spaces (for the background and terminology see any
textbook on probability, such as~\cite{probability}, and measure
theory~\cite{halmos}):
\par
The source ensemble $\cE$ is described by a measure space $\Omega$
(with probability measure $P$), and a measurable map
$\varphi:\Omega\longrightarrow \cP(\cH)\subset\cS(\cH)$ from
$\Omega$ into the set of pure states on the Hilbert space $\cH$
(which is still of finite dimension $d$), mapping
$\omega\in\Omega$ to $\proj{\varphi_\omega}$. We can then easily
define encoding and decoding $(E,D)$ for blocks of length $n$:
\begin{eqnarray}
\label{eq:measurable:E} E: \Omega^n
&\longrightarrow& \cS(\cH_B) \times \Omega_C, \\
\label{eq:measurable:D} D: \cB(\cH_B)\otimes\cB(\ell^2(\Omega_C))
&\longrightarrow& \cB_d^{\ox n},
\end{eqnarray}
where $E$ is a Markov kernel, $\Omega_C$ is a \emph{finite set}, and
$D$ is CPTP. The quantification of classical and quantum resources we
adopt unchanged, and the fidelity condition reads as follows: the
combined encoding and decoding gives rise to a Markov kernel
\begin{equation}
D\circ E: \Omega^n \longrightarrow \cB_d^{\ox n},
\end{equation}
and, using the abbreviation
\begin{equation}
(D\circ E)(\omega_1\ldots\omega_n)
              =\int_{\cB(\cH_B)} (D\circ E)({\rm d}\sigma|\omega_1\ldots\omega_n) \sigma,
\end{equation}
we require that
\begin{equation}
\label{eq:infinite:fidelity} F=\int_{\Omega^n} P^{\otimes n}({\rm
d}\omega_1\ldots\omega_n)
      F\bigl(\varphi_{\omega_1\ldots\omega_n},(D\circ E)(\omega_1\ldots\omega_n)\bigr)
                                                                                \geq 1-\e.
\end{equation}
\par
Let us denote by $\mu$ the measure induced by $P$ and this Markov kernel on
$\Omega\times\cS(\cH_B)\times\Omega_C$:
\begin{equation}
\mu(F_A\times G_{BC})
:= \int_{F_A} P({\rm d}\omega)E(G_{BC}|\omega).
\end{equation}
We denote
its restrictions (marginals) to factors $\Omega_A=\Omega$,
$\cS(\cH_B)$, $\Omega_C$ by $P=\mu_A$, $\mu_B$, $q:=\mu_C$, respectively, and
analogously $\mu_{AC}$, etc.
\par
With the help of Radon--Nikodym derivatives we can always construct
the Bayesian ``inverse'' Markov kernel
\begin{equation}
q: \Omega_C \longrightarrow \Omega_A\times\cS(\cH_B)
\end{equation}
that gives rise to the same joint
distribution:
\begin{equation}
\int_{G_C} \mu_C({\rm d}j)q(F_{AB}|j) = \mu(F_{AB}\times G_C).
\end{equation}
In fact, $\mu_C$--almost everywhere,
\begin{equation}
q(F_{AB}|j)=\frac{{\rm d}\mu(F_{AB}\times\{j\})}{{\rm d}\mu_C(j)}.
\end{equation}
\par
To follow the procedure of section~\ref{sec:lowerBound} we have to
define the relevant information quantities (for their properties,
see~\cite{Gray,OhyaP}):
\par
First, $S(A:C)$ can be expressed as $D( \mu_{AC}\| \mu_A\ox\mu_C )$,
in terms of the relative entropy (or Kullback--Leibler divergence) of two
measures
\begin{equation}
\label{eq:divergence} D(\mu\|\lambda):=\int \mu({\rm
d}x)\log\left(\frac{{\rm d}\mu(x)}{{\rm d}\lambda(x)}\right),
\end{equation}
where $\frac{{\rm d}\mu(x)}{{\rm d}\lambda(x)}$ denotes the
Radon--Nikodym derivative. If this does not exist $\mu$--almost
everywhere, we define $D(\mu\|\lambda)=\infty$. It is a fact that
in Eq.~(\ref{eq:divergence}) the Radon--Nikodym derivative always
exists, and it can be checked that in the finite case the new
definition coincides with the old.
\par
Second, $S(A:B|C)=\int_{\Omega_C} q({\rm d}j) S(A:B|C=j)$, with
$S(A:B|C=j)$ denoting the quantum mutual information associated to
the conditional probability measure $q(\cdot|j)$ on
$\Omega_A\times\cS(\cH_B)$: for any such distribution $\lambda$, with
first marginal $\lambda_A$ and Markov kernel
$L:\Omega_A\rightarrow\cS(\cH)$,
\begin{equation}
S_\lambda(A:B)=S\left(\int_{\cS(\cH)}\lambda_B({\rm d}\sigma)\sigma\right)
                  -\int_{\Omega_A} \lambda_A({\rm d}\omega)
                   S\left(\int_{\cS(\cH)}L({\rm d}\sigma|\omega)\sigma\right).
\end{equation}
Again, it is possible to check that for discrete probability spaces
we obtain the same expressions as before.
\par
The proofs of lemmas~\ref{lemma:Mproperties}
and~\ref{lemma:additivity} and of theorem~\ref{thm:lowerbound} are
directly adapted to this language, essentially replacing all sums
representing probability averages by integrals. (Note that even
the ``continuity in $\e$'' part in the latter applies as the
functions $f$ and $g$ depend only on $\e$ and $d$.) This is
possible since the monotonicity and convexity properties we used
are still true in the infinite setting.
\par
At the end of the proof we arrive at encodings mapping
$\omega\in\Omega$ to $\proj{\varphi_\omega}\otimes\sum_j
p(j|\omega)\proj{j}$ (i.e., the corresponding Markov kernel maps $i$
to the point mass at $\proj{\varphi_\omega}$ times a discrete measure
on $\Omega_C$). Such encodings we denote
``$p:\Omega_A\rightarrow\Omega_C$'', and we get
\begin{equation}
Q^*(R) \geq
\inf_{p:\Omega_A\rightarrow\Omega_C,\ |\Omega_C|<\infty}
                                            \{ S(A:B|C) : S(A:C)\leq R \}.
\end{equation}
\par
Dropping the finiteness of $\Omega_C$ can only decrease the lower
bound, and we arrive the following general version of
theorem~\ref{thm:lowerbound}:
\begin{theorem}
\label{thm:lowerbound:infinite} For the ensemble
$\cE=(\Omega,P,\varphi)$, $$Q^*(R) \geq
M(\cE,R):=\inf_{p:\Omega_A\rightarrow\Omega_C}
                                            \{ S(A:B|C) : S(A:C)\leq R \},$$
with
\begin{eqnarray*}
  S(A:C)   &=& D(\mu\|P\ox q), \\
  S(A:B|C) &=& \int_{\Omega_C} q({\rm d}j)
                  S\left(\int_{\Omega_A}q({\rm d}\omega|j)\proj{\varphi_\omega}\right),
\end{eqnarray*}
where $\mu$ is the measure on $\Omega_A\times\Omega_C$ induced by $P$
and the Markov kernel $p(\cdot | \cdot)$, $q$ is its marginal on
$\Omega_C$ and $q(\cdot|\cdot)$ is the Bayesian Markov kernel
$\Omega_C\rightarrow\Omega_A$. \qed
\end{theorem}

\subsection{Adaptation of the coding theorem}
\label{subsec:infinite:coding} The obstacles to an application of our
coding scheme, proposition~\ref{prop:coding}, are the potentially
infinite range of the source register ($\Omega$) and the classical
encoding ($\Omega_C$). Of course, when in the previous subsection
we allowed the latter to be infinite, we only made $M$ smaller,
and at that point it was not clear that this was a good move.
\par
The purpose of the present subsection is to show that it is
possible to approximate the effect of an infinite encoding by a
strictly finite one: finitely many possible states on ${\cal H}$
and finitely many classical symbols. This will inevitably
introduce some error, that we'll have to counter by a suitably
adapted notion of typical subspace.
\begin{lemma}
  \label{lemma:partition}
  For $\epsilon>0$ there exists a partition of $\cS(\cH)$ into
  $m\leq C(d)\epsilon^{-d^2}$ Borel sets each of which has radius
  at most $\epsilon$: in each part $\cS_i$ there exists a
  state $\sigma_i$ such that for all $\rho\in\cS_i$,
  $\| \rho-\sigma_i \|_1 \leq \e$.
  The constant $C(d)$ depends only on $d$.
\end{lemma}
\begin{proof}
The set of states on $\cH$ is affinely isomorphic to the set of
positive complex $d\times d$--matrices with trace $1$, which is
contained in the set of selfadjoint complex matrices with all
$d^2$ real and imaginary parts of entries in the interval
$[-1,1]$: this is a $d^2$--dimensional hypercube. This can be
partitioned into $(2\sqrt{2}d^3)^{d^2}\e^{-d^2}$ many small
hypercubes of edge length $\e/(d^3\sqrt{2})$. It is easy to check
that for any $\rho,\sigma$ in the same small cube,
$\|\rho-\sigma\|_1 \leq \e$.
\end{proof}
\par
For a source $(\Omega,P,\varphi)$ such a partition entails a
partition $\cZ$ of $\Omega$ into at most $m$ measurable pieces
$Z_i$, with $\omega_i\in Z_i$ such that
$\proj{\varphi_{\omega_i}}=\sigma_i$. (We need only consider
pieces that intersect the image of $\varphi$.) A central role will
be played by the ``contraction'' of the infinite ensemble $\cE$ to
the finite ensemble
$\cE'=\{\varphi_{\omega_i},\widehat{P}(i)=P(Z_i)\}$ which is
obtained by identifying all of $Z_t$ to the single state
$\varphi_{\omega_i}$.
\par
We have already defined the set of $\widehat{P}$--typical sequences
$\cT_{\widehat{P},\d}$, and now can define the following typical set
for $P$:
\begin{equation}
\label{eq:typical}
  \cT^{\cZ}_{P,\d}:= \bigcup_{I\in\cT_{\widehat{P},\d}} Z_{i_1}\times\cdots\times Z_{i_n}.
\end{equation}
It obviously inherits the large probability property of
$\cT_{p',\d}$:
\begin{equation}
\label{eq:Z:typical:probability}
  P^{\otimes n}\bigl( \cT^{\cZ}_{P,\d} \bigr) \geq 1-\frac{1}{\d^2}
\end{equation}
\par\medskip
Before we can describe the coding scheme we have to introduce a
variant of the conditional typical sequences and subspaces: for a
channel $W:\cI\rightarrow\cJ$ and $\d,\e>0$ define
\begin{equation}
\label{eq:conditional:typical:eps} \cT^{(\e)}_{W,\d}(I) := \bigl\{ J:
\forall ij\ |N(ij|IJ)-N(i|I)W(j|i)|
                                             \leq \d\sqrt{N(i|I)}+\e N(i|I)\bigr\}.
\end{equation}
(Our previous notion is recovered with $\e=0$, and in the sequel
$\e$ will be small, compared to $\d$ which we shall choose large.)
Observe that this is a union of conditional type classes. Using
Eq.~(\ref{eq:conditional:typical:upper}) it is quite easy to show
that
\begin{equation}\begin{split}
\label{eq:conditional:typical:eps:card} |\cT^{(\e)}_{W,\d}(I)| &\leq
(n+1)^{|\cI||\cJ|}
           \exp\left( nH(W|P_I)+\sum_i N(i|I)|\cJ|\eta\bigl(\e+\d N(i|I)^{-1/2}\bigr) \right) \\
                       &\leq (n+1)^{|\cI||\cJ|}
           \exp\left( nH(W|P_I)+ n|\cJ|\eta(\e)+n\eta(\d|\cI|/\sqrt{n}) \right),
\end{split}\end{equation}
where we have used the inequality $\eta(x+y) \leq \eta(x)+\eta(y)$
and concavity of $\eta$.
\par
Similarly, for a collection of states $W_i$, which we endow with
fixed diagonalizations $W_i=\sum_{j=1}^d W(j|i)\proj{e_{j|i}}$, we
can define the projector
\begin{equation}
\label{eq:conditional:typical:proj:eps} \Pi^{(\e)}_{W,\d}(I) :=
\sum_{J\in\cT^{(\e)}_{W,\d}(I)} \proj{e_{J|I}},
\end{equation}
and get from Eq.~(\ref{eq:conditional:typical:eps:card}) the estimate
\begin{equation}
\label{eq:conditional:typical:proj:eps:card} \Tr\Pi^{(\e)}_{W,\d}(I)
\leq  (n+1)^{d|\cI|}
                             \exp\left( nH(W|P_I)+ nd\eta(\e)+n\eta(\d|\cI|/\sqrt{n}) \right).
\end{equation}
Its other most important property that we shall use is the following:
consider a product state $\sigma=\sigma_1\ox\cdots\ox\sigma_n$ such
that, with some $I=i_1\ldots i_n$,
\begin{equation}
\label{eq:epsilon:close:averages} \forall i\qquad \left\|
\frac{1}{N(i|I)}\sum_{k:i_k=i}\sigma_k-W_i \right\|_1 \leq \e.
\end{equation}
Then we claim that
\begin{equation}
\label{eq:robust:weak:law:proj}
\Tr\bigl(\sigma\Pi^{(\e)}_{W,\d}(I)\bigr) \geq 1-\frac{|\cI|}{\d^2}.
\end{equation}
The proof goes as follows: the left hand side above does not change if
we replace $\sigma_k$ by $\sigma_k' :=\sum_j
\proj{e_{j|i_k}}\sigma_k\proj{e_{j|i_k}}$, because the projector is a
sum of one--dimensional projectors $\proj{e_{J|I}}$. Thus we may
assume that $\sigma_k$ has diagonal form in the chosen eigenbasis of
$W_{i_k}$: $\sigma_k=\sum_j S_k(j)\proj{e_{j|i_k}}$.
\par
Note that the left hand side of Eq.~(\ref{eq:robust:weak:law:proj})
can be rewritten as $(S_1\ox\cdots\ox
S_n)\bigl(\cT^{(\e)}_{W,\d}(I)\bigr)$, a classical probability. Now
it is immediate from the definition of the latter set
(Eq.~(\ref{eq:conditional:typical:eps})) and from the
condition~(\ref{eq:epsilon:close:averages}) on $\sigma$ that
\begin{equation}
\cT^{(\e)}_{W,\d}(I) \supset \cT_{\overline{S},\d}(I),
\end{equation} with the
channel $\overline{S}(j|i)=\frac{1}{N(i|I)}\sum_{k:i_k=i}S_k(j)$.
Hence
\begin{equation}\begin{split}
  (S_1\ox\cdots\ox S_n)\bigl(\cT^{(\e)}_{W,\d}(I)\bigr)
         &\geq (S_1\ox\cdots\ox S_n)\bigl(\cT_{\overline{S},\d}(I)\bigr)        \\
         &\geq \left(1-\frac{1}{\d^2}\right)^{|\cI|} \geq 1-\frac{|\cI|}{\d^2},
\end{split}\end{equation}
the second line by Chebyshev's inequality.
\par
After these preparations we are ready to prove the infinite source
version of proposition~\ref{prop:coding}:
\begin{proposition}
\label{prop:infinite:coding} Let $\cE=(\Omega_a,P,\varphi)$ be a
source. For a probability distribution $P$ on $\Omega$ and a Markov
kernel $p(\cdot|\cdot):\Omega_A\rightarrow\Omega_C$, $\e>0$, there
exists a partition $\cZ$ of $\Omega_A$ into $m-1<C(d)\e^{-d^2}$
measurable sets, corresponding to an $\e$--fine partition of the
state space, and for $\d>0$ a visible code $(E,D)$ such that
$$\forall\omega=(\omega_1\ldots\omega_n)\in\cT^{\cZ}_{P,\delta}\qquad
      F\bigl( \proj{\varphi_\omega},(D\circ E)(\omega) \bigr) \geq 1-\frac{4m^2}{\delta^2}.$$
and sending
\begin{eqnarray*}
  nS(A:C)  + nKm^2\eta(\d/\sqrt{n}) +K'm^2\log(n+1)           & & \text{classical bits}, \\
  nS(A:B|C)+ n\bigl(3dm^2\eta(2\delta m^2/\sqrt{n})+3d\eta(\e)\bigr) + dm\log(n+1)
                                                              & & \text{quantum bits}.
\end{eqnarray*}
\end{proposition}
\begin{proof}
  We can find the partition by lemma~\ref{lemma:partition} and the
  discussion thereafter.
  \par
  Consider now the (measurable) coarse--graining map
  \begin{equation}
  T:\omega \longmapsto i\in\{1,\ldots,m-1\}\text{ for }\omega\in Z_i.
  \end{equation}
  Applying $T$ to $\Omega_A$ (and the identity map to $\cB(\cH_B)$ and $\Omega_C$)
  leads to a new distribution $\mu'$ on $\Omega_{A'}\times\cB(\cH_B)\times\Omega_C$,
  with $\Omega_{A'}=\{1,\ldots,m-1\}$. By the data--processing
  inequality~\cite{csiszar:koerner,OhyaP} we have
  \begin{equation}
  \label{eq:monotonicity:1}
  S(A':C) \leq S(A:C)\text{ and }S(A':B|C) \leq S(A:B|C).
  \end{equation}
  \par
  Next we change the quantum part of the encoding by collecting all the
  weight of a piece $Z_i$ into $\varphi_i:=\varphi_{\omega_i}$: we can do
  this by a similar coarse--graining map
  \begin{equation}
  \widetilde{T}:\sigma \longmapsto \proj{\varphi_i}\text{ for }\sigma\in Z_i.
  \end{equation}
  The resulting distribution will be denoted by $\mu''$: it is supported
  on a finite set $\Omega_{A'}$ and a finite set of states $\varphi_i$
  (in fact, the ``contracted'' ensemble $\cE'$ of the discussion after
  lemma~\ref{lemma:partition}). It is generated by a Markov kernel
  $\hat{p}:\Omega_{A'}\rightarrow\Omega_C$, which in this case is simply a
  finite collection of (conditional) distributions $\hat{p}(\cdot|i)$ on
  $\Omega_C$. Note that this is a valid encoding in the sense of the
  definition of $M(\cE',R)$, in the main section.
  Let us denote the corresponding conditional quantum mutual information
  by $S(A':B'|C)$.
  \par
  By definition of $S(A':B|C)$ and the partition $\cZ$, we have
  \begin{equation}
  \label{eq:monotonicity:2}
  S(A':B'|C) \leq S(A':B|C) + 2d\eta(\e/d),
  \end{equation}
  using Fannes' inequality~(\ref{eq:fannes}) twice.
  \par
  To end this step--by--step discretization, we may change the
  encoding to a stochastic matrix
  $p':\Omega_{A'}\rightarrow\{1,\ldots,m\}=:\Omega_{C'}$, by the
  considerations of section~\ref{sec:lowerBound} (see also
  proposition~\ref{prop:small:classical:part}), such that
  \begin{equation}
  \label{eq:monotonicity:3}
  S(A':B'|C')\leq S(A':B'|C)\text{ and }S(A':C')=S(A':C).
  \end{equation}
  \par
  So finally, we are in a position to apply the coding method of
  proposition~\ref{prop:coding}, with the sole difference that
  we use for the quantum encoding the projector $\Pi^{(\e)}_{p',\d}(I)$
  instead of our previous conditional typical projector,
  and $I$ is such that $\omega_1\ldots\omega_n\in Z_I$.
  \par
  The fidelity estimate is obtained just like there, only using
  Eq.~(\ref{eq:robust:weak:law:proj}).
  The classical rate estimate we copy from proposition~\ref{prop:coding},
  and for the quantum rate estimate, we follow its derivation
  in the proof, using Eq.~(\ref{eq:conditional:typical:proj:eps:card})
  to estimate the range of the projectors $\Pi^{(\e)}_{p',\d}(I)$:
  we have to send
  \begin{equation}
  nS(A':B'|C')+ n\bigl(3dm^2\eta(2\delta m^2/\sqrt{n})+d\eta(\e)\bigr) + dm\log(n+1)
  \end{equation}
  quantum bits, which, by Eqs.~(\ref{eq:monotonicity:1})--(\ref{eq:monotonicity:3}),
  yields our desired estimate.
\end{proof}
\par\medskip
This immediately leads to the result that we wanted:
\begin{theorem}
\label{thm:infinite:coding}
  For any ensemble $\cE=(\Omega,P,\varphi)$,
  $$Q^*(R)=M(\cE,R).$$
\end{theorem}
\begin{proof}
  That $M(\cE,R)$ is a lower bound to $Q^*$ is proved
  by theorem~\ref{thm:lowerbound:infinite}. For its achievability
  choose $\e>0$ and a Markov kernel $p$ such that both $S(A:C)\leq R$ and
  $S(A:B|C)\leq M(\cE,R)+\e$.
  \par
  Choose now a partition $\cZ$ according to
  proposition~\ref{prop:infinite:coding}, fixing $m$. Now choose $\d$
  large enough, so that according to that proposition a code exists
  which has fidelity $1-\e$ on a state set of probability $1-\e$,
  i.e., it has average fidelity $1-2\e$ on the ensemble.
  By the proposition it has cbit rate $S(A:C)+o(1)$ and qubit rate
  \begin{equation}
  S(A:B|C)+2\eta(\e)+o(1)\leq M(\cE,R)+2\eta(\e)+\e+o(1),
  \end{equation}
  as $n\rightarrow\infty$. As $\e$ was arbitrary, our claim is proved.
\end{proof}

\subsection{On the AVS in the infinite setting}
\label{subsec:infinite:AVS} With the help of the above
proposition~\ref{prop:infinite:coding} the case of an arbitarily
varying source of an \emph{infinite} ensemble is dealt with easily,
in much the same way as we did in the finite case (see
section~\ref{sec:avs}):
\par
Formally, of course, an arbitrarily varying source is a triple
$(\Omega,{\bf P},\varphi)$, where $\Omega$ and $\varphi$ are a
measurable space and a measurable map into states, as before, and
${\bf P}$ is a set of probability distributions on $\Omega$.
\par
With the definitions of encoding and decoding from
subsection~\ref{subsec:measure} we require
\begin{equation}
\label{eq:infinite:avs:fidelity} \forall P^n\in{\bf P}^n\quad
 \int_{\Omega^n} P^{\otimes n}({\rm d}\omega_1\ldots\omega_n)
                        F\bigl(\proj{\varphi_\omega},(D\circ E)(\omega)\bigr) \geq 1-\e.
\end{equation}
Denoting the trade--off function as $Q^*(R,{\bf P})$, we obtain
the expected result:
\begin{theorem}
\label{thm:infinite:avs} $Q^*(R,{\bf P})=M({\bf P},R)$, with $$M({\bf
P},R) = \sup_{P\in{\bf Q}} M(P,R),$$ where ${\bf Q}={\rm conv}({\bf
P})$ is the convex hull of ${\bf P}$.
\end{theorem}
\begin{proof}
  The inequality ``$\geq$'' is obvious, like in the finite case: the adversary
  can certainly always mock up an i.i.d.~source $P\in{\bf Q}$, hence
  theorem~\ref{thm:lowerbound:infinite} applies.
  \par
  For the opposite inequality, we start by choosing an $\e>0$ and a partition $\cZ$
  according to proposition~\ref{prop:infinite:coding}. Every distribution $P$
  in ${\bf P}$ gives rise to a distribution $\widehat{P}\in\cP_{m-1}$, and
  we denote
  \begin{equation}
  \widehat{\bf P}:=\big\{ \widehat{P} : P\in{\bf P} \bigr\}.
  \end{equation}
  Note that, because the map $P\mapsto\widehat{P}$ is affine linear, we get
  $\widehat{\bf Q}={\rm conv}(\widehat{\bf P})$.
  \par
  Now for $\d>0$ we introduce again the set
  \begin{equation}
  \cT := \bigcup_{\widehat{P}\in\widehat{\bf Q}} \cT_{\widehat{P},\d},
  \end{equation}
  and it is easy to see (compare Eq.~(\ref{eq:Z:typical:probability})) that
  \begin{equation}
  \cT^{\cZ} := \bigcup_{I\in\cT} Z_{i_1}\times\cdots\times Z_{i_n}
  \end{equation}
  carries $1-\d^{-2}$ of the probability of every $P^n\in{\bf P}^n$.
  On the other hand, because $\cT$ is a union of type classes,
  we can find ``few'' $\widehat{P}_1,\ldots,\widehat{P}_T$,
  $T\leq (n+1)^m$ such that the corresponding $\cT_{\widehat{P}_t,\d}$ cover
  $\cT$.
  The coding is very simple: on seeing a state $\varphi_{\omega_1\ldots\omega_n}$
  the encoder finds the index $I$ of the piece $Z_I$ in the partition $\cZ^n$
  such that $\omega_1\ldots\omega_n\in Z_I$,
  and the type of $I$. If $I\in\cT$, he looks up $t$ such that
  $I\in\cT_{\widehat{P}_t,\d}$ and uses the coding scheme of
  proposition~\ref{prop:infinite:coding} for $\widehat{P}_t$.
  (Note that he needs not even send the type of $I$ as that is part
  of the protocol of proposition~\ref{prop:infinite:coding}.)
  Choosing $\delta$ large enough this recipe gives a code with high fidelity
  for every $P^n\in{\bf P}^n$; by construction and
  proposition~\ref{prop:infinite:coding}, it has rates of
  $R+o(1)$ cbits and $M({\bf P},R)+f(\e)+o(1)$ qubits, with a function
  $f(\e)$ that tends to $0$ as $\e\rightarrow 0$.
\end{proof}
\par\medskip
To end this discussion, we would like to point out that a similar
treatment of remote state preparation can be done: in fact, as we
discussed in section~\ref{sec:rsp}, we always use the ``$1$ ebit $+$
$1$ cbit per qubit'' technique (theorem~\ref{thm:1and1per1}) on top
of an efficient trade--off coding. To do this for an infinite
ensemble one only has to understand that the bound of
theorem~\ref{thm:1and1per1} is strong enough to allow approximation
of the set of projected (compressed) product states
$\varphi_{\omega_1}\ox\cdots\ox\varphi_{\omega_n}$, at negligible
additional classical cost.

\section{Discussion and conclusions}
Our main result is a simple formula for the trade--off between
quantum and classical resources in visible compression.  The
formula expresses the trade--off curve $Q^*(R)$ in terms of a
single--letter optimization over conditional probability
distributions of bounded size. This unexpectedly simple resolution
places optimal trade--off coding into a small but growing class of
problems in quantum information theory whose answers are not only
known in principle but can be calculated in practice. (Another
notable recent addition is the entanglement--assisted capacity of
a quantum channel \cite{bennett:et:al}.)

At a conceptual level, for any given ensemble $\cE$ of quantum
states, $Q^*(R)$ can be thought of as a quantitative description
of how ``classical'' the ensemble is.  Any deviation from
classicality is captured in the trade--off curve in the form of
inefficiency of the classical storage. The amount of information
that can be extracted from many copies of $\cE$ while causing
negligible disturbance, for example, can be read directly off the
curve by identifying the point at which classical resources begin
to become inefficient as compared to quantum.  Much more subtle
indicators of classicality are also available in $Q^*(R)$,
however. We saw, for instance, that for the parameterized BB84
ensemble, $Q^*(R)$ had a kink at the point corresponding to
partitioning the ensemble into nearly orthogonal subensembles.

Going beyond the compression of ensembles, we saw that it is
possible to formulate a version of our main result in the setting
of arbitrarily varying sources, corresponding to the situation in
which the encoder and decoder have only partial or even no
knowledge of the distribution of input states.  Despite this
handicap, compression is frequently still possible and we once
again find that the trade--off curve can be calculated via a
tractable optimization problem.  For ensembles with symmetry, the
problem can even often be reduced to calculating $Q^*(R)$ for one
particular ensemble.  Thus, for any given set of pure states,
including the whole manifold of states on a given Hilbert space,
these tools allow us to calculate the rate of exchange from qubit
storage to classical storage.  The answer is given, of course, not
in terms of a single number but as the trade--off curve. (Like in
any market, the going rate depends on supply.)

Our view that $Q^*(R)$ encodes the balance of quantum and
classical information in a given ensemble or set of states is
further bolstered by the role it was found to play in optimal
remote state preparation.  In this context, the minimal amount of
classical communication required for any given rate of
entanglement consumption can, once again, be read directly off the
quantum--classical trade--off curve.  That the comparatively
exotic process of remote state preparation should reduce, via
theorem \ref{thm:1and1per1}, to visible compression is a
tremendous simplification.

Of course, while we have seen that the results of this paper
resolve some basic questions about trading different types of
resources in quantum information, most related questions remain
open. To begin, it is possible to trade entanglement, quantum
communication and classical communication all together in a
generalized type of remote state preparation. Since our results
here describe the two extremes when first entanglement and then
quantum communication are not permitted, it seems likely that
similar techniques could resolve the full trade--off ``surface''.
More ambitiously, one could define channel capacities for noisy
quantum channels that interpolate between the fully quantum and
classical capacities by studying the usefulness of a channel for
simultaneously sending quantum and classical information.  The
problem analogous to the trade--off question studied here would be
to determine the achievable \emph{region} of quantum--classical
rate pairs.  Unfortunately, given that neither the fully classical
nor fully quantum extremes are fully understood, it may be a long
time before we develop tools capable of analyzing that problem.

Therefore, to end, we offer two related open problems that are
perhaps closer to the realm of the tractable. First, it would be
useful to have a set of rules for extracting qualitative
features of the trade--off curve, such as the location of any kinks
and perhaps more detailed differentiability properties, 
from the structure of the input
states (or ensemble). Second, it would be an interesting challenge
to apply the observations of section \ref{sec:symmetry} on
symmetry to the explicit calculation of the trade--off curve for
particular examples and, more generally, to find other approaches
to simplifying these calculations.

%Second, is there any direct relationship between the trade--off
%curve for a given ensemble and its accessible information
%\cite{Fuchs}?

\section*{Acknowledgments}
We thank Charles H. Bennett, David P. DiVincenzo, Daniel Gottesman,
Debbie W. Leung, Michael A. Nielsen, D\'enes Petz, Peter W. Shor and John. A.
Smolin for enlightening discussions and helpful suggestions.
\par
P.H. was supported by US National Science Foundation grant no.~EIA--0086038
and a Sherman Fairchild Fellowship.
R.J. and A.W. are supported by the U.K.~Engineering and Physical
Sciences Research Council.

\appendix
\section{Proofs of auxiliary propositions}

\subsection{Proof of proposition \ref{prop:useEquality}}
\label{pfprop:useEquality}
\begin{proof}
Suppose the classical
register $C$ decomposes into parts $C_1$ and $C_2$ with corresponding
joint density operator
\begin{eqnarray}
\r^{A B C_1 C_2} = \sum_i p_i \proj{i}^A \ox \proj{\ph_i}^B \ox
\sum_{j,k} p(i|j,k) \proj{j}^{C_1} \ox \proj{k}^{C_2}.
\end{eqnarray}
If we define the conditional ensembles $\cE_{jk}$ and $\cE_j$, then
\begin{eqnarray}
S(A:B|C_1 C_2) = \sum_{jk} q_{jk} S(\cE_{jk}) \leq S(A:B|C_1) =
\sum_j q_j S(\cE_j)
\end{eqnarray}
by the concavity of the von Neumann entropy.
\par
Therefore, for any map with $S(A:C_1) < R \leq H(p)$, we can always
adjoin a second classical register $C_2$ such that $S(A:C_1 C_2)=R$
without increasing the conditional mutual information.
\end{proof}

\subsection{Proof of proposition \ref{prop3.two}} \label{pfprop3.two}
\begin{proof}
W.l.o.g.~let $i\in\{1,\ldots,m\}$. The information quantities in the
definition of $M$ can be re--expressed as follows:
\begin{align}
  S(A:B|C) &=\sum_j q_j S\left(\sum_i q(i|j)\proj{\varphi_i}\right), \\
  S(A:C)   &=H(p)-\sum_j q_j H\bigl(q(\cdot|j)\bigr),
\end{align}
with $q_j=\sum_i p_i p(j|i)$ and $q_j q(i|j)=p_i p(j|i)$. We read $q$
as a probability distribution on the set ${\cal P}_m$ of all
probability distributions on $\{1,\ldots,m\}$. Thus the minimization
problem in the definition of $M$ can be expressed as finding the
infimum of $\sum_j q_j S\bigl(f(q(\cdot|j))\bigr)$ over the set
$${\cal P}(p,R)=\left\{ q\text{ p.d. on }{\cal P}_m:\sum_j q_j
q(\cdot|j)=p,
                              \sum_j q_j H\bigl(q(\cdot|j)\bigr)\geq H(p)-R \right\},$$
where $f$ is an affine linear function on probability distributions,
mapping the distribution $p$ to the quantum state $\sum_i
p_i\proj{\varphi_i}$.
\par
Now we argue structurally: the set ${\cal P}(p,R)$ is convex (as a
subset of an infinite dimensional probability simplex with additional
linear inequality constraints), and the aim function is linear. Hence
the infimum is an infimum over the extreme points of ${\cal P}(p,R)$,
which are, by Caratheodory's theorem, distributions $q$ with support
at most $m+1$, the number of inequalities that define ${\cal
P}(p,R)\subset{\cal P}({\cal P}_m)$, see e.g.~\cite{convexity}. In
section~\ref{sec:symmetry} proposition
\ref{prop:small:classical:part} we will give a detailed exposition of
a more general form of this result.
\end{proof}

\subsection{Proof of proposition \ref{prop:tensorM}}\label{appendix:tensorM}
\begin{proof}
 The ``$\leq$'' inequality follows directly by forming the tensor product of
  two encodings for $\cE_1$
and $\cE_2$
 with classical rates $R_1$ and $R_2$ respectively.\\
The ``$\geq$'' inequality is shown by choosing an encoding for the
tensor product with classical rate
 $R$ and then using the chain rule several times for subdivisions
 $A=A_1A_2$ and $B=B_1B_2$ as follows. First observe that
 \begin{equation}
  R \geq S(A_1A_2:C) = S(A_1:C)+S(A_2:C|A_1)
                  =: R_1   + R_2
 \end{equation}
 and then
 \begin{eqnarray}
 S(A_1A_2:B_1B_2|C) & = &
 S(A_1:B_1B_2|C)+S(A_2:B_1B_2|C,A_1)\\
           &     \geq & S(A_1:B_1|C)  +S(A_2:B_2|C,A_1) \nonumber\\
              &  \geq & M(\cE_1,R_1)+\inf\{S(A_2:B_2|C,A_1) :
              S(A_2:C|A_1)\leq
              R_2 \} \nonumber \\
              &  \geq  & M(\cE_1,R_1)+M(\cE_2,R_2) \nonumber \\
          &      \geq & \min \{M(\cE_1,R_1)+M(\cE_2,R_2) : R_1+R_2=R \} \nonumber
 \end{eqnarray}
 The second last line is seen as follows: in the line
 above it, the two mutual informations are conditional on $A_1$, so they both
 can be written as averages over the values of $A_1$. Hence the inequality follows
 by the convexity of $M$ in $R$.
 \end{proof}

\subsection{Proof of proposition \ref{prop:SchurConcave}}
\label{appendix:SchurConcave}
\begin{proof}  It is sufficient to verify that any encoding operator
\begin{eqnarray}
\r^{ABC} = \sum_{ik} p_i a_k \proj{i}^A \ox \proj{k}^A \ox U_k
\proj{\ph_i} U_k^{\dg B} \ox \sum_j p(j|i,k) \proj{j}^C
\end{eqnarray}
for $\cF$ gives rise to a valid encoding operator
\begin{eqnarray}
\s^{ABC} = \sum_i p_i \proj{i}^A \ox \proj{\ph_i}^B \ox \sum_{jk}
p(j|i,k) a_k \proj{j}^C \ox \proj{k}^C
\end{eqnarray}
for $\cE$ satisfying $S_\s(A:B|C) \leq S_\r(A:B|C)$ and $S_\s(A:C)
\leq S_\r(A:C)$.
\end{proof}

\subsection{Proof of proposition \ref{prop:equality:conditions}}
\label{appendix:equality:conditions}
\begin{proof}
We will first prove the proposition for irreducible $\cE$.
Using a trick introduced by Holevo \cite{Holevo73}, we can
reduce the problem further to the case of a two--state ensemble:
for an ensemble $\{\r_i^B \ox \s_i^C,p_i\}$ of states (we assume that
all $p_i>0$) and two specific indices $k$ and $l$, define a new index
\begin{equation}
  j(i):=\begin{cases}
             i & i\neq k,l,   \\
             * & i\in\{k,l\}.
        \end{cases}
\end{equation}
(Of course, in the case we have in mind, the $\rho_i$ are the pure states
from the ensemble $\cE$, and the $\sigma_i$ are commuting
mixed states representing the classical information.)
Then consider the multipartite state
$$\Omega=\sum_i p_i \proj{i}^{A_1}\ox\proj{j(i)}^{A_2}\ox\r_i^B\ox\s_i^C.$$
The definition of $j(i)$ and the familiar chain rule imply
\begin{equation}
  \label{eq:chain}
  S(A_1:BC)=S(A_1 A_2:BC)=S(A_2:BC)+S(A_1:BC|A_2).
\end{equation}
Note that the second term is an average over the
values of $j(i)$ of Holevo quantities for the corresponding
reduced ensembles.  Therefore, it has only one nonzero contribution,
which is
\begin{equation}
S(A_1:BC|A_2)=(p_k+p_l)\chi\bigl(\{\r_i\ox\s_i,p_i/(p_k+p_l)\}_{i=k,l}\bigr).
\end{equation}
Then, using Eq.~(\ref{eq:chain}) and monotonicity of $\chi$ under partial
trace repeatedly:
\begin{equation}\begin{split}
  \chi\bigl(\{p_i,\rho_i\otimes\sigma_i\}\bigr)
             &=    S(A_1:BC) = S(A_2:BC)+S(A_1:BC|A_2)      \\
             &\geq S(A_2:B)
                  +(p_k+p_l)
                   \chi\bigl(\{\rho_i\otimes\sigma_i,p_i/(p_k+p_l)\}_{i=k,l}\bigr) \nonumber \\
             &\geq S(A_2:B)
                  +(p_k+p_l)\chi\bigl(\{\rho_i,p_i/(p_k+p_l)\}_{i=k,l}\bigr)
                  \nonumber \\
             &=    S(A_2:B)+S(A_1:B|A_2)=S(A_1:B) \nonumber \\
             &=    \chi\bigl(\{\rho_i,p_i\}\bigr). \nonumber
\end{split}\end{equation}
Assuming that the first and the last Holevo quantity have the same value,
we must have equality in the third line, implying
\begin{equation}
  \label{eq:central1}
  \chi\bigl(\{\rho_i\otimes\sigma_i,q_i\}_{i=k,l}\bigr)
                 = \chi\bigl(\{\rho_i,q_i\}_{i=k,l}\bigr),
\end{equation}
with $q_i=p_i/(p_k+p_l)$.
Then, applying the general formula
\begin{equation}
\chi\bigl(\{\omega_i,p_i\}\bigr) = \sum_i p_i D(\omega_i\|\omega),
\end{equation}
to Eq.~(\ref{eq:central1}),
with $\omega=\sum_i p_i\omega_i$ and $D$ the relative entropy function,
and using Lindblad monotonicity once more
yields
\begin{equation}
  \label{eq:central2}
  D(\rho_k\otimes\sigma_k\|q_k\rho_k\otimes\sigma_k+q_l\rho_l\otimes\sigma_l)
                                                = D(\rho_k\|q_k\rho_k+q_l\rho_l).
\end{equation}
(And likewise for $l$.)
\par
With this we are almost done: invoking a result of Ohya and Petz (see
Ref.~\cite{OhyaP}, theorem 9.12) we conclude that there exists a
CPTP map $R$ such that
\begin{align}
  R(\rho_k)              &= \rho_k\otimes\sigma_k, \\
  R(q_k\rho_k+q_l\rho_l) &= q_k\rho_k\otimes\sigma_k+q_l\rho_l\otimes\sigma_l,
\end{align}
from which it follows by linearity that
\begin{equation}
R(\rho_l) = \rho_l\otimes\sigma_l.
\end{equation}
Since CPTP maps ($R$ and $\Tr_C$) cannot decrease fidelity we thus must
have $\rho_k\perp\rho_l$ or $\sigma_k=\sigma_l$.
\par
In the particular case that the initial ensemble is irreducible
we conclude that all $\sigma_i$ must be equal, or else the partial
trace over $C$ strictly decreases the Holevo quantity.
If the ensemble $\cE$ is not irreducible, a simple variation on
the previous argument shows that, for each of the irreducible
subensembles $\cE_l$, $\chi(\cE_l)$ must be equal to $\chi$ of the
corresponding subensemble $\{\ph_{il}\ox\s_{il},p_{i|l}\}$ of $\cF^{BC}$.
Applying our conclusions to these subensembles finishes the proof of
the proposition.
\end{proof}

\subsection{Proof of proposition \ref{prop:small:classical:part}
\label{pfprop:small}}
\begin{proof}
As explained earlier in the proof of proposition \ref{prop3.two}, any
classical encoding map can be viewed as a probability distribution
$q$ on the set $\cP_{\cI}$ of probability distributions on $\cI$ with
barycenter $p$: $p=\sum_j q_j q(\cdot|j)$.
\par
Covariance of the encoding means invariance of $q$ under the natural
action of $G$ on $\cP_{\cI}$, i.e., $g:p \longmapsto p^g$. Hence for
each distribution $p$ in the support of $q$ we must have all the
$p^g$ in the support as well. On the other hand, we need far less
conditions to obey, as it will turn out:
\par
Assume that the covariant encoding is given by the distributions
$$\bigl(q(\cdot|j)\bigr)^g\text{ with probability }\frac{1}{|G|}q_j,\
g\in G,j=1,\ldots .$$ Now choose representatives $i_1,\ldots,i_t$ of
the orbits, and observe that (by $G$--invariance)
\begin{equation}
\sum_{j,g} \frac{1}{|G|}q_j \bigl(q(\cdot|j)\bigr)^g = p
\end{equation}
if and only if
\begin{equation}
\label{eq:sum:to:p} \forall \tau=1,\ldots,t\quad
    \sum_{j,g} \frac{1}{|G|}q_j q\bigl(g^{-1}i_\tau|j\bigr) = p(i_\tau).
\end{equation}
Similarly, $S(A:C)\leq R$ if and only if
\begin{equation}
\label{eq:entropy:bounded} \sum_j q_j H\big(q(\cdot|j)\bigr) \geq
H(p)-R,
\end{equation}
and finally, our aim function reads
\begin{equation}
\label{eq:ABC:expression} S(A:B|C)=\sum_{j,g} \frac{1}{|G|}q_j
S\left( \sum_i q(i|j)\proj{\varphi_{gi}} \right).
\end{equation}
\par
Now consider the affine linear map from $\cP_{\cI}$ to $\RR^{t+1}$
defined by
\begin{equation}
A : p \longmapsto
        \left( H(p) ; \frac{1}{|G|}\sum_g p(g^{-1}i_\tau) : {\tau=1,\ldots,t} \right).
\end{equation}
Note that the image of this map is in a certain $t$--dimensional
subspace because, if $t-1$ of the conditions~(\ref{eq:sum:to:p})
are satisfied then the $t^{\text{th}}$ is also, automatically.
Eqs.~(\ref{eq:sum:to:p}) and~(\ref{eq:entropy:bounded}) are really
conditions on the $q_j$--weighted average of the the images
$A_j=A\bigl(q(\cdot|j)\bigr)$, $A=\sum_j q_j A_j$. By
Caratheodory's theorem~\cite{convexity} the same average can be
obtained by convex combination of $t+1$ of these i.e. by a
distribution $q'$ on the $j$'s with support containing at most
$t+1$ points. In fact, $q$ is easily seen to be expressible as a
convex combination of such small support distributions, say
$q^{\prime (a)}$ with weights $\lambda_a$.
\par
To conclude, we observe that our aim function in
Eq.~(\ref{eq:ABC:expression}) is \emph{linear} in the distribution
$q$: hence, it is the $\lambda_a$--weighted sum of similar such
expressions with $q^{\prime (a)}$ in place of $q$. For one value of
$a$ at least this is smaller than $S(A:B|C)$, the corresponding
$q^{\prime (a)}$ satisfies $\sum_j q^{\prime (a)} A_j=A$, and hence
Eqs.~(\ref{eq:sum:to:p}) and~(\ref{eq:entropy:bounded}). As explained
in the remark preceding the statement of proposition
\ref{prop:small:classical:part}, to obtain a $G$--covariant encoding
we can split up each $q(\cdot|j)$ (with $j$ in the support of
$q^{\prime (a)}$) into the $G$ translated distributions
$\bigl(q(\cdot|j)\bigr)^g$, proving the claim.
\par
%In fact, we have proved even slightly more, namely that $\cJ$ can be
%assumed to have at most $t+1$ orbits under the $G$--action.
\end{proof}

\end{document}